\documentclass[a4paper, thm-restate]{lipics-v2021}
\nolinenumbers
\hideLIPIcs

\usepackage[utf8]{inputenc}

\newcommand{\anoncite}[1]{\ifdefined\anon\cite{#1:anon}\else\cite{#1:full}\fi}

\usepackage{nicefrac}
\usepackage{tikz}
\usetikzlibrary{shapes}
\usepackage{mathtools}
\usepackage{tabularx}
\usepackage{booktabs}
\usepackage{bm}
\usepackage{graphicx}

\DeclareMathOperator*{\argmax}{arg\,max}

\usepackage{hyphenat}
\usepackage[ruled,vlined,linesnumbered]{algorithm2e} 
\SetAlFnt{\footnotesize}
\usepackage{float}
\renewcommand{\SetProgSty}[1]{\renewcommand{\ProgSty}[1]{\textnormal{\csname#1\endcsname{##1}}\unskip}}%
\SetKwProg{Fn}{function}{ is}{end}
\SetKwFor{With}{with}{do}{end}

\SetProgSty{myargstyle}
\SetArgSty{myargstyle}
\SetKwComment{tcc}{// }{}

\SetCommentSty{mycommentstyle}

\def\tsconst{TS/const}

\title{Tailstorm: A Secure and Fair Blockchain for Cash Transactions}

\ifdefined\anon
\author{Anonymous Authors}{\strut}{}{}{}
\authorrunning{Anonymous Authors}
\Copyright{Anonymous Authors}
\else
\Copyright{Patrik Keller, Ben Glickenhaus, George Bissias, and Gregory Griffith}
\authorrunning{P. Keller, B. Glickenhaus, G. Bissias, and G. Griffith}
\author{Patrik Keller}{Universität Innsbruck}{tailstorm.on.arxiv@pkel.dev}{}{}
\author{Ben Glickenhaus}{University of Massachusetts Amherst}{}{}{}
\author{George Bissias}{University of Massachusetts Amherst}{}{}{}
\author{Gregory Griffith}{Bitcoin Unlimited}{}{}{}
\fi


\begin{document}

\maketitle

\begin{abstract}

  Proof-of-work (PoW) cryptocurrencies rely on a balance of security and fairness in order to maintain a sustainable ecosystem of miners and users.
  Users demand fast and consistent transaction confirmation, and in exchange drive the adoption and valuation of the cryptocurrency.
  Miners provide the confirmations, however, they primarily seek rewards.
  In unfair systems, miners can amplify their rewards by consolidating mining power.
  Centralization however, undermines the security guarantees of the system and might discourage users.

  In this paper we present Tailstorm, a cryptocurrency that strikes this balance.
  Tailstorm merges multiple recent protocol improvements addressing security, confirmation latency, and throughput with a novel incentive mechanism improving fairness.
  We implement a parallel proof-of-work consensus mechanism with $k$~PoWs per block to obtain state-of-the-art consistency guarantees~\cite{keller2022ParallelProofofwork}.
  Inspired by Bobtail~\cite{bissias2020BobtailImproved} and Storm~\cite{awemany2019StormWhitepaper}, we structure the individual PoWs in a tree which, by including a list of transactions with each PoW, reduces confirmation latency and improves throughput.
  Our proposed incentive mechanism discounts rewards based on the depth of this tree.
  Thereby, it effectively punishes information withholding, the core attack strategy used to reap an unfair share of rewards.

  We back our claims with a comprehensive analysis.
  We present a generic system model which allows us to specify Bitcoin, $\mathcal B_k$\,\cite{keller2022ParallelProofofwork}, and Tailstorm from a joint set of assumptions.
  We provide an analytical bound for the fairness of Tailstorm and Bitcoin in \emph{honest} networks and we confirm the results through simulation.
  We evaluate the effectiveness of \emph{dishonest} behaviour through reinforcement learning.
  Our attack search reproduces known optimal strategies against Bitcoin, uncovers new ones against $\mathcal B_k$, and confirms that Tailstorm's reward discounting makes it more resilient to incentive layer attacks.
  Our results are reproducible with the material provided online~\anoncite{cpr:repo}.

  Lastly, we have implemented a prototype of the Tailstorm cryptocurrency as a fork of Bitcoin Cash.
  The client software is ready for testnet deployment and we also publish its source online~\anoncite{clientImpl}.
\end{abstract}

\section{Introduction} \label{sec:introduction}

Proof-of-work (PoW) cryptocurrencies can be thought of as stacked systems comprising four layers.
The PoW layer moderates access of weakly identified parties using a mining puzzle: proposing a new block requires finding a small hash.
The consensus layer allows all participants to agree on a specific block ordering.
The application layer maintains a distributed ledger by writing cryptocurrency transactions into the blocks.
Finally, the incentive layer motivates participation in the PoW layer by minting new cryptocurrency for successful miners.

The circular dependencies between the layers result in interdependent failures.
Consensus faults are inevitable when individual miners gain too much control~\cite{garay2020SoKConsensus}.
Unreliable consensus enables double spending and erodes confidence in the cryptocurrency.
Devaluation of the currency renders the mining rewards worthless as well.
Lastly, misaligned incentives encourage centralization among the miners, eventually allowing the strongest one to break consensus.

In this paper, we introduce and analyze Tailstorm, a new cryptocurrency that strengthens two layers of the stack. We employ an innovative incentive mechanism as well as a state-of-the-art consensus mechanism, while retaining the PoW and application layers of Bitcoin.
We draw inspiration from the organization of delta blocks in the Storm protocol~\cite{awemany2019StormWhitepaper} as well as the use of partial PoW in the Bobtail~\cite{bissias2020BobtailImproved} and $\mathcal{B}_k$ protocols~\cite{keller2022ParallelProofofwork}.

Bitcoin's consensus~\cite{nakamoto2008BitcoinPeertopeer} uses a \emph{sequential} PoW mechanism where each block references a single parent block.
The blocks form a tree and the participants mine new blocks that extend the longest branch according to the \emph{longest chain rule}.
Blocks off the longest branch are discarded.
Evidently, attackers who possess more than 50\,\% of the hash rate pose a threat to the system:
they can execute double-spend attacks by mining their own branch until it eventually becomes the longest.
But even less mining power might suffice since honest participants also discard the blocks of other honest miners if they are not on the same branch.
This happens naturally in realistic networks due to propagation delays.
If an attacker can induce and exploit communication delays, all discarded blocks may benefit the attacker; effectively increasing their strength~\cite{guo2022BitcoinLatency}.

In contrast, Tailstorm implements a \emph{parallel} PoW consensus mechanism that largely avoids discarding blocks.
For this, we closely follow the approach taken by Keller and Böhme at AFT\,'22~\cite{keller2022ParallelProofofwork}. Their protocol, $\mathcal B_k$, confirms each block with $k$ \emph{votes}.
$\mathcal B_k$ blocks do not require a PoW, only votes do.
Notably, votes confirming the same block can be mined in parallel because they do not depend on each other.
Discarding only occurs when there are more than $k$ votes for the same block.
Even then, individual discarded votes only account for $\nicefrac1k$ of a block's PoW.
As Keller and Böhme~\cite{keller2022ParallelProofofwork} argue, this makes consensus more robust against consensus layer attacks.

But it is futile to analyze the consensus without considering incentives.
Ideally, rewards are distributed fairly, which means that a miner's expected reward is proportional to its hash rate.
Arnosti and Weinberg~\cite{arnosti2022BitcoinNatural} show that even small inequalities in reward allocation encourage substantial centralization of hash rate which ultimately poses a threat to consensus and the cryptocurrency itself~\cite{garay2020SoKConsensus}.

\begin{figure}
  \centering
  \begin{tikzpicture}[
  >=stealth,
  x=\linewidth/62,
  y=1ex,
  subb/.style={draw, circle},
  summary/.style={draw, rectangle, inner sep=1.5ex},
  lbl/.style={font=\footnotesize, align=left},
  align=center,
  ]
  \small
  \node[summary] (s0)  at ( 0, 0) {};
  \node[subb]    (w0)  at ( 7,-2) {};
  \node[subb]    (w1)  at ( 5, 2) {};
  \node[subb]    (w2)  at ( 9, 2) {};
  \node[summary] (s1)  at (14, 0) {};
  \node[subb]    (w3)  at (19, 0) {};
  \node[subb]    (w4)  at (23, 0) {};
  \node[subb]    (w5)  at (27, 0) {};
  \node[summary] (s2)  at (32, 0) {};
  \node[subb]    (w6)  at (38,-3) {};
  \node[subb]    (w7)  at (38, 0) {};
  \node[subb]    (w8)  at (38, 3) {};
  \node[summary] (s3)  at (44, 0) {};
  \node[subb]    (w9)  at (49, 0) {};
  \node[subb]    (w10) at (53,-2) {};
  \node[subb]    (w11) at (53, 2) {};
  \node[summary] (s4)  at (58, 0) {};
  \draw[->] (w0) -- (s0);

  \draw[->] (w0) -- (s0);
  \draw[->] (w1) -- (s0);
  \draw[->] (w2) -- (w1);
  \draw[->] (s1) -- (w0);
  \draw[->] (s1) -- (w2);

  \draw[->] (w3) -- (s1);
  \draw[->] (w4) -- (w3);
  \draw[->] (w5) -- (w4);
  \draw[->] (s2) -- (w5);

  \draw[->] (w6) -- (s2);
  \draw[->] (w7) -- (s2);
  \draw[->] (w8) -- (s2);
  \draw[->] (s3) -- (w6);
  \draw[->] (s3) -- (w7);
  \draw[->] (s3) -- (w8);

  \draw[->] (w9) -- (s3);
  \draw[->] (w10) -- (w9);
  \draw[->] (w11) -- (w9);
  \draw[->] (s4) -- (w10);
  \draw[->] (s4) -- (w11);

  \newcommand{\lbl}[1]{
    \def\vdepth{#1}
    $\begin{aligned}
      d &= \vdepth \\
      r &= \nicefrac{\vdepth}{3}
    \end{aligned}$
  }

  \draw[dashed] ([shift={(-2.5, 5)}] s1) rectangle ([shift={(3, -11)}] s0);
  \draw[dashed] ([shift={(-2.5, 5)}] s2) rectangle ([shift={(3, -11)}] s1);
  \draw[dashed] ([shift={(-2.5, 5)}] s3) rectangle ([shift={(3, -11)}] s2);
  \draw[dashed] ([shift={(-2.5, 5)}] s4) rectangle ([shift={(3, -11)}] s3);

  \path (s1) -- node[shift={(0.25,-8)}, midway, lbl] {\lbl{2}} (s0);
  \path (s2) -- node[shift={(0.25,-8)}, midway, lbl] {\lbl{3}} (s1);
  \path (s3) -- node[shift={(0.25,-8)}, midway, lbl] {\lbl{1}} (s2);
  \path (s4) -- node[shift={(0.25,-8)}, midway, lbl] {\lbl{2}} (s3);
\end{tikzpicture}
  \caption{
    Example of a Tailstorm blockchain with $k = 3$ subblocks per summary.
    Squares are summary blocks, circles are subblocks, arrows indicate hash-references.
    Dashed rectangles mark subblock trees with depth~$d$ and discounted subblock reward $r$.
    Reduced depth implies lower rewards.
  }
  \label{fig:tailstormchain}
\end{figure}
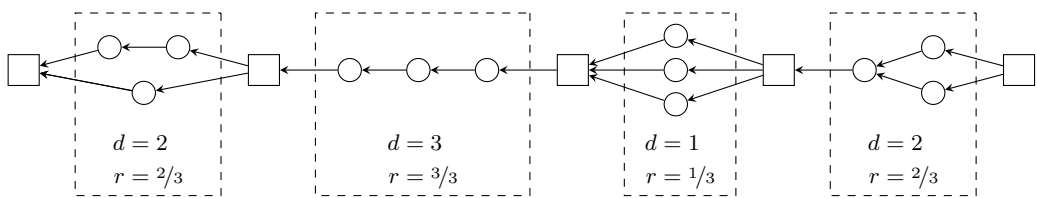

Unfairness arises from natural network delays and dishonest behavior.
Both of these factors affect Bitcoin.
In latent networks, two blocks mined around the same time may refer to the same parent, even if both miners follow the longest chain rule.
One of the blocks will be discarded, the other rewarded.
The stronger miner has more hash rate to support their own block, hence the weaker miner is worse off.
Network-level attackers can additionally exploit latency to manipulate impartial participants in their favor.
But even without delays, Bitcoin miners with more than one third of the hash rate can reap an unfair share of rewards by temporally withholding blocks instead of acting honestly~\cite{eyal2014MajorityNot, gervais2016SecurityPerformance, negy2020SelfishMining, sapirshtein2016OptimalSelfish}.
As we will demonstrate, $\mathcal B_k$ suffers from the same problem, partly due to its leader election mechanism.

Tailstorm addresses the unfairness problem through an innovative reward scheme that punishes withholding.
The Tailstorm blockchain consists of \emph{subblocks} and \emph{summary} blocks.
Similar to $\mathcal B_k$, summaries do not require a PoW, but subblocks do.
Assembling a new summary requires $k$~subblocks that confirm the same parent summary.
To preserve the security properties of $\mathcal B_k$, subblocks confirming the same summary are conflict-free, and hence can be mined in parallel.
Taking inspiration from Bobtail~\cite{bissias2020BobtailImproved}, subblocks \emph{optionally} refer to another subblock instead of a summary, and hence form a tree.
With Tailstorm, we propose to discount rewards based on the depth of this tree, as depicted in Figure~\ref{fig:tailstormchain}.
Mining subblocks in private causes branching of the tree, reduces its depth, and ultimately leads to lower rewards.

We support our claims with a comprehensive analysis and make the following contributions.
\begin{enumerate}
  \item
    We formulate a generic system model for PoW cryptocurrencies.
    The models abstracts PoW and communication, defines a joint set of assumptions, and enables valid comparisons between different consensus protocols and incentive mechanisms.
  \item
    We specify Bitcoin, $\mathcal B_k$, and Tailstorm in the joint model.
    To isolate the effect of Tailstorm's discount reward scheme and $\mathcal B_k$'s leader election mechanism, we additionally specify a hybrid protocol, \tsconst, modelling Tailstorm without discounting and $\mathcal B_k$ without leader election.
  \item
    We provide an upper bound for the orphan rate of Tailstorm in honest networks.
    Compared to Bitcoin~\cite{rizun2016SubchainsTechnique}, Tailstorm creates less orphans and hence presents less opportunity for unfairness.
  \item
    We implement the system model as a simulator and show that Tailstorm is more fair than Bitcoin in honest but realistic networks with propagation delays.
    We confirm Bitcoin's inherent tradeoff: while short block intervals are desirable for fast confirmations and a less volatile stream of rewards, they also bias rewards in favour of strong miners.
    In Tailstorm, these concerns are largely separated by configuring long \emph{summary} block intervals for fairness and short \emph{subblock} intervals for fast confirmations and frequent rewards.
  \item
    We evaluate multiple hard-coded attack strategies against the specified protocols, finding that attacks which are profitable against $\mathcal B_k$ are less profitable against Tailstorm, with the \tsconst{} protocol lying in between.
  \item
    We follow Hou et al.~\cite{hou2021SquirRLAutomating} and search optimal attack strategies using reinforcement learning.
    Our search reproduces optimal strategies against Bitcoin~\cite{gervais2016SecurityPerformance,sapirshtein2016OptimalSelfish} and generally matches or outperforms the hard-coded strategies.
    The regularity of our results indicates that we indeed found near-optimal strategies against all protocols and enables the conclusion that Tailstorm is less susceptible to incentive layer attacks than the other protocols.
  \item
    We describe the Tailstorm application layer, which implements a cryptocurrency on top of the proposed consensus protocol.
    It preserves Bitcoin's transaction logic while enabling faster confirmations.
    Transactions are stored in subblocks in much the same way that the Storm protocol~\cite{awemany2019StormWhitepaper} stores transactions in delta blocks.
  \item
    Lastly, we implement a prototype of Tailstorm which is ready for testnet deployments and make the code available online~\anoncite{clientImpl}.
\end{enumerate}

We structure the paper in order of our contributions.
Section~\ref{sec:model} defines the system model and Section~\ref{sec:spec} presents the specification of Tailstorm; specification of the remaining protocols is deferred to Appendix~\ref{apx:protocols}.
In Section~\ref{sec:fairness}, we evaluate the protocols in an honest network with propagation delays.
In Section~\ref{sec:attacks}, we evaluate hard-coded attack strategies, and in Section~\ref{sec:reinforcement_learning} we conduct the search for optimal policies with reinforcement learning.
Section~\ref{sec:implementation} presents the Tailstorm cryptocurrency and our prototype implementation.
In Section~\ref{sec:discussion}, we discuss related work, limitations and future work.
Section~\ref{sec:conclusion} concludes.

\section{System Model} \label{sec:model}

Recall the layered view on PoW cryptocurrencies presented in the introduction:
The PoW layer moderates access using a mining puzzle,
the consensus layer establishes a specific block ordering,
the application layer writes cryptocurrency transactions into the blocks,
and the incentive layer mints new cryptocurrency for successful miners.
We now present a system model that abstracts PoW and communication to enable concise specification of the consensus layer. Application and incentives are considered in later sections.

In practice, PoW consensus protocols are executed as distributed systems, where independent nodes communicate over a P2P network.
Messages exchanged between the nodes may be subject to natural or potentially malicious delays.
To facilitate specification, we abstract this distributed system and model it algorithmically.
We define a virtual environment that emulates distributed protocol execution in a single thread of computation.
Within the virtual environment, nodes are represented as numbers, and blocks are represented as vertices in a directed acyclic graph (DAG).
Mining is simulated as a loop with random delays, while communication is modeled by restricting the visibility of blocks to a subset of the nodes.
The behaviour of nodes is defined by functions, which can be customized to model different protocols.

We define the virtual environment in Algorithm~\ref{alg:env}.
The environment maintains a DAG where each \emph{vertex} represents one block.
Each block $b$ has an associated list of parent blocks that constitute the outgoing \emph{edges} in the DAG.
We denote this list $\texttt{parents}(b)$.
In practice, edges arise from hash references pointing to other blocks in the blockchain.
We say that block~$b$ is a \emph{descendant} of block~$b'$, if~$b'$ is either a parent of~$b$, or is connected transitively by the $\texttt{parents}$ relationship.
In this case, we say~$b'$ is an \emph{ancestor} of~$b$.
Each block has \emph{properties} which are assigned as the protocol unfolds.
For example, the virtual environment uses the Boolean property $\texttt{pow}(b)$ to track whether block $b$ has a PoW or not.

\begin{algorithm}[t]
  \caption{Virtual Environment}
  \label{alg:env}

  \BlankLine
  $t \gets \texttt{Root}()$\tcc*[r]{\texttt{Root}: protocol's genesis block template}
  $b \gets $ block obtained from appending template $t$ to the block DAG\tcc*[r]{reification}
  \For(\tcc*[f]{$n$: number of nodes}){$i = 1, \dots, n$}{
    $\texttt{tip}(i) \gets b$\tcc*[r]{$\texttt{tip}(i)$: state of node $i$}
    $\texttt{visibility}(b, i) \gets$ true\label{l:env:vis1}\tcc*[r]{$\texttt{visibility}(b, i)$: whether node $i$ sees block $b$}
  }

  \BlankLine
  \While(\tcc*[f]{concurrent PoW loop}){true}{\label{l:powloopa}
    sample delay $d_\text{pow} \sim \texttt{Expon}(\text{rate}\,\lambda)$\tcc*[r]{$\lambda$: PoW rate}
    sample node $i \sim \texttt{Discrete}(1,\dots,n\ \text{with weights}\,\kappa_1, \dots, \kappa_n)$\label{l:powminer}%
    \tcc*[r]{$\kappa_1, \dots \kappa_n$: relative hash rates}
    wait for $d_\text{pow}$ seconds, handling other tasks concurrently\;
    \With(\tcc*[f]{partial and immutable view on block DAG}){local view of node $i$\label{l:env:view1}}{
      $\emph{tmpl} \gets \texttt{Extend}(\texttt{tip}(i))$\tcc*[r]{\texttt{Extend}: protocol's mining rule}
    }
    $b \gets $ block obtained from appending template \emph{tmpl} to the block DAG\tcc*[r]{reification}
    $\texttt{pow}(b') \gets$ true\tcc*[r]{mark block as mined}
    $\texttt{visibility}(b, \cdot) \gets$ false\tcc*[r]{block not yet propagated}
    \lIf(\tcc*[f]{\texttt{Validate}: protocol's chain structure}){$\texttt{Validate}(b)$}{%
      $\texttt{Deliver}(b, i)$%
    }\label{l:powloopb}
  }

  \BlankLine
  \Fn(\tcc*[f]{Turn block visible for node \dots}){$\texttt{Deliver}(\text{block } b, \text{node } i)$}{
    \For(\tcc*[f]{\dots{} in topological order \dots}){$p \in \texttt{parents}(b)$}{
      wait until $\texttt{visibility}(p, i)$ is true\;
    }

    \If(\tcc*[f]{\dots{} and at most once.}){not $\texttt{visibility}(b,i)$}{
      $\texttt{visibility}(b,i) \gets$ true\label{l:env:vis2}\;

      \BlankLine
      \With(\tcc*[f]{partial and immutable view on block DAG}){local view of node $i$\label{l:env:view2}}{
        ($\texttt{tip}(i)$, \textit{share}, \textit{append}) $\gets \texttt{Update}(\texttt{tip}(i), b)$%
        \tcc*[r]{\texttt{Update}: protocol's state update rule}
      }

      \BlankLine
      \For(\tcc*[f]{handle block broadcast requests}){$b' \in$ \textit{share} and $j =1,\dots,n$, $j \neq i$}{
        pick message delay $d_\text{net}$ according to network assumptions\;
        $\texttt{Deliver}(b', i)$ in $d_\text{net}$ seconds\;
      }

      \BlankLine
      \For(\tcc*[f]{handle requests to append blocks without PoW}){$\emph{tmpl} \in$ \textit{append}}{
        $b' \gets $ block obtained from appending template $\emph{tmpl}$ to the block DAG\tcc*[r]{reification}
        $\texttt{pow}(b') \gets$ false\tcc*[r]{block has no PoW}
        $\texttt{visibility}(b', \cdot) \gets$ false\;
        \lIf(\tcc*[f]{\texttt{Validate}: protocol's chain structure}){$\texttt{Validate}(b')$}{%
          $\texttt{Deliver}(b', i)$%
        }
      }
    }
  }
\end{algorithm}

We label the participating nodes with integers ranging from~$1$ to~$n$.
For each node, the virtual environment maintains a local view of the DAG and a preferred tip of the chain.
In Lines~\ref{l:env:view1} and~\ref{l:env:view2}, we restrict the local view of node~$i$ to blocks where $\texttt{visibility}(b,i)$ was set to true.
Initially, local views are empty and new blocks are not visible to any node.
We denote as $\texttt{tip}(i)$ the preferred \emph{tip} of node~$i$, the block to which a new block from~$i$ will point.
We describe the behaviour of nodes as pure functions.
These functions are called by the environment to obtain instructions from the node which the environment then interprets according to our assumptions.
This makes all modifications of the DAG and all communication explicit in Algorithm~\ref{alg:env}.
In particular, nodes do not directly append blocks to the DAG; they return block templates, which the environment then reifies by appending a new block to the DAG.

A protocol is fully specified through four functions:
\texttt{Root} and \texttt{Validate} define the structure of the blockchain, while \texttt{Update} and \texttt{Extend} define the behavior of honest nodes.

The \texttt{Root} function takes no argument and returns a single block, which we call \emph{genesis}.
Initially, the genesis is the only block in the DAG.
\texttt{Validate} takes a block as argument and returns \texttt{true} if the block is valid and \texttt{false} otherwise.
E.\,g., Bitcoin's \texttt{Validate} checks the \texttt{pow} property and that there is exactly one parent.
The environment enforces block validity during the reification of blocks, while deployed protocols would reject invalid blocks in the communication layer.
The genesis is not subject to the validity rule.

The \texttt{Update} function specifies how nodes react to newly visible blocks, after they are mined locally with PoW, appended locally without PoW, or received from the network.
The function takes two arguments: the node's currently preferred tip and the new block.
The function returns the new preferred tip, a list of blocks the node intends to \emph{share} with other nodes, and a list of block templates it wants to \emph{append} to the chain \emph{without} PoW.
On the other hand, the function \texttt{Extend} defines how nodes grow the chain \emph{with} PoW.
It takes a single argument, a node's currently preferred tip, and returns a template for the block that the node intends to mine.

We follow related work~\cite{sompolinsky2015SecureHighrate,rizun2016SubchainsTechnique, dembo2020EverythingRace, li2021CloseLatency, keller2022ParallelProofofwork} and model the mining process in continuous time.
The virtual environment generates independent mining delays from the exponential distribution $\texttt{Expon}(\lambda)$, with rate $\lambda$ measured in expected number of proofs-of-work per second.
Accordingly, the expected value of the distribution, $\nicefrac1\lambda$, is called the \emph{mining interval} and is measured in seconds.
After each mining delay, the environment randomly selects a successful miner, obtains a block template from \texttt{Extend}, and reifies the block by appending it to the DAG.
We support arbitrary hash rate distributions among the nodes by setting the weights $\kappa_1, \dots, \kappa_n$ in Line~\ref{l:powminer} accordingly.

The \texttt{Deliver} function captures the process of making blocks visible to nodes.
The specified protocols have in common that block validation requires knowledge of all referenced blocks.
We avoid a lot of boilerplate code in the specification, by ensuring that parent blocks are delivered before their children.
Upon delivery, the virtual environment first invokes the \texttt{Update} function to obtain the node's new preferred tip, a list of blocks the node wants to share, and a list of block templates the node intends to append without PoW.
It then handles the node's requests to share and append.
Communication is modelled through delayed delivery, while appends happen immediately.

\section{The Tailstorm Protocol} \label{sec:spec} \label{sec:tailstorm_protocol}

This section specifies the Tailstorm consensus protocol and reward mechanism using the algorithmic model described in Section~\ref{sec:model}.
The specification serves as the basis for our theoretical analyses, network simulations, and attack search in  subsequent sections.
We first describe Tailstorm's chain structure in Section~\ref{sec:spec:chain}.
We then specify the behaviour of honest nodes in Section~\ref{sec:spec:node}.
As a point of reference, we also specify Bitcoin and $\mathcal B_k$ protocols in Appendix~\ref{apx:protocols}.
In this section, we assume that the application layer implements a cryptocurrency which we can use to pay  rewards.
We defer the description of Tailstorm's application layer to Section~\ref{sec:implementation}.
Throughout this section, we focus on honest miners who follow the protocol as intended.
Later sections will consider dishonest behaviour.

\subsection{Chain Structure} \label{sec:spec:chain}

\begin{algorithm}[t]
  \caption{Tailstorm: Chain Structure}
  \label{alg:ts:valid}

  \Fn{$\texttt{Root}()$}{
    \Return block template $b$ with $\texttt{summary}(b) = \texttt{true}$ and
    $\texttt{height}(b) = \texttt{depth}(b) = 0$\;
  }

  \BlankLine
  \Fn{$\texttt{Validate}(\text{block } b)$}{
    \uIf{$\texttt{summary}(b)$}{
      $p \gets$ last summary before $b$\;
      $S \gets$ subblocks between $b$ and $p$\;
      \Return $(|S| = k) \land (\texttt{depth}(b) = 0) \land
      (\texttt{height}(b) = \texttt{height}(p) + 1)$\;
    }
    \Else(\tcc*[f]{$b$ is subblock}){
      $p \gets$ first parent of $b$\;
      \Return $ \texttt{pow}(b) \land (|\texttt{parents}(b)| = 1) \land
        \left(\texttt{depth}(b) = \texttt{depth}(p) + 1\right) \land
        \left(\texttt{height}(b) = \texttt{height}(p)\right)$\;
    }
  }
\end{algorithm}

Algorithm~\ref{alg:ts:valid} defines Tailstorm's chain structure.
Each block is either a summary or a subblock.
Subblocks must have PoW and they must have exactly one parent.
Summaries do not require PoW and reference $k$ subblocks, each confirming the same ancestor summary.

Each block $b$ has two integer properties, $\texttt{height}(b)$ and $\texttt{depth}(b)$.
The genesis has height and depth zero.
Subblocks inherit the height of their parent and they increment the depth by one.
Summaries increment the height and reset the depth to zero.
Figure~\ref{fig:tailstormchain} in Section~\ref{sec:introduction} illustrates a valid Tailstorm chain for $k=3$.
Note that the subblocks confirming the same summary form a tree and that the $\texttt{depth}$ property tracks the depth of this tree.
The $\texttt{height}$ property on the other hand counts the number of trees that have been summarized.

To incentivize participation, Tailstorm allocates rewards to the miners of subblocks.
The reward size is proportional to the depth of the subblock tree:
let $b$ be a summary block, and $S$ be the set of subblocks in the corresponding subblock tree.
Then all subblocks in $S$ are allocated the same reward
\begin{align}
  \texttt{discount}(b) = \frac ck \cdot \max_{x \in S}(\texttt{depth}(x) )\,,\label{f:discount}
\end{align}
where $c$ represents a tunable upper limit on the subblock reward.
In Figure~\ref{fig:tailstormchain} we show the rewards for $c = 1$.
Note that the reward scheme punishes non-linearities in the blockchain, and this punishment affects all included subblocks equally.

\subsection{Honest Nodes}\label{sec:spec:node}

\begin{algorithm}[t]
  \caption{Tailstorm: Node}
  \label{alg:ts:node}

  \Fn{\label{l:ts:node:pref:b}$\texttt{Preference}(\text{summary } s, \text{summary } b)$}{
    \emph{hs}, \emph{hb} $\gets \texttt{height}(s), \texttt{height}(b)$\;
    \emph{ns}, \emph{nb} $\gets$ size of subblock tree confirming $s$ and $b$\;
    \emph{rs}, \emph{rb} $\gets$ reward obtained from $(s, b)$\;
    \Return $(hb > hs) ~\lor
      (hb = hs \land nb > ns) ~\lor
      (hb = hs \land nb = ns \land rb > rs)$\label{l:ts:node:pref:e}\;
  }

  \BlankLine
  \Fn(\tcc*[f]{$t$: currently preferred block}){\label{l:ts:node:extend:b}$\texttt{Extend}(\text{tip } t)$}{
    \leIf{$t$ has children}{$p \gets$ maximum-depth subblock after $t$}{$p \gets t$}
    \Return subblock template $b$ with\\
    $\quad\texttt{parents}(b) = [p]$,
    $\texttt{height}(b) = \texttt{height}(p)$, and
    $\texttt{depth}(b) = \texttt{depth}(p) + 1$\label{l:ts:node:extend:e}\;
  }

  \BlankLine
  \Fn(\tcc*[f]{$t$: currently preferred block // $b$: new block}){\label{l:ts:node:upd:b}$\texttt{Update}(\text{tip } t, \text{block } b)$}{
    $\emph{pref}, \emph{share}, \emph{append} \gets t, [b], []$\label{l:ts:node:share}%
    \tcc*[r]{new preferred tip, blocks to broadcast, templates to append}
    \uIf{$\texttt{summary}(b)$}{
      \lIf{$\texttt{Preference}(\emph{pref},b)$}{$\emph{pref} \gets b$\label{l:ts:node:updsum1}}
    }
    \Else(\tcc*[f]{$b$ is subblock}){
      $p \gets$ last summary before $b$\;
      \lIf{$\texttt{Preference}(\emph{pref},p)$}{$\emph{pref} \gets p$\label{l:ts:node:updsum2}}
      $S \gets$ all subblocks in subblock tree confirming $p$\;
      \If{$|S| \geq k$\label{l:ts:node:next}}{
        $L \gets$ result of Algorithm~\ref{alg:subblock_selection}: Subblock Selection\;
        create summary template $s$ with $\texttt{parents}(s) = L$\;
        \emph{append} $\gets [s]$\;
      }
    }
    \Return \emph{pref}, \emph{share}, \emph{append}\label{l:ts:node:upd:e}\;
  }
\end{algorithm}

Algorithm~\ref{alg:ts:node} specifies the behaviour of honest nodes.
The algorithm revolves around a preference order (ln.\,\ref{l:ts:node:pref:b}-\ref{l:ts:node:pref:e}) that ranks summaries first by height, then by number of confirming subblocks, and finally by potential personal reward for the individual node.
Nodes set the highest ranked summary as their preferred summary (ln.\,\ref{l:ts:node:updsum1}+\ref{l:ts:node:updsum2}) and they mine subblocks (ln.\,\ref{l:ts:node:extend:b}-\ref{l:ts:node:extend:e}) that confirm their preferred summary.
To maximize the depth of the subblock tree, nodes append their subblocks to the longest existing branch.

Whenever nodes learn about a new block (ln.\,\ref{l:ts:node:upd:b}-\ref{l:ts:node:upd:e}), they share it with the other nodes and they update their preference.
As soon as there are $k$ subblocks (ln.\,\ref{l:ts:node:next}) confirming the preferred summary, nodes assemble the next summary.
When there are more than $k$ subblock candidates for the next summary, nodes choose the subblocks to maximize their own rewards.
We present a greedy algorithm for subblock selection in Algorithm~\ref{alg:subblock_selection}.

\begin{algorithm}[t]
  \caption{Tailstorm: Subblock Selection}
  \label{alg:subblock_selection}
  $R \gets \emptyset$\tcc*{selected subblocks}
  \While(\tcc*[f]{select one subblock per iteration}){$|R| < k$}{
    $C \gets S \setminus R$\tcc*{$S$: candidate subblocks}
    \For{$x \in C$}{
      $B_x \gets x$ and all ancestors of $x$ in $S$\;
      $B_x' \gets B_x \setminus R$\tcc*{newly referenced blocks}
      $r_x \gets$ number of node's own subblocks in $B_x'$\tcc*{reward}
    }
    $C \gets \{ x \in C : |R| + |B_x'| \leq k \}$\tcc*{enforce tree size $k$}
    $y \gets \argmax_{x \in C}{r_x}$\tcc*{select candidate with highest reward}
    $R \gets R \cup B_y'$\tcc*{track referenced subblocks}
  }
  \Return leaves in subblock tree $R$\tcc*{invariant: $|R| = k$}
\end{algorithm}

\subsection{Difficulty Adjustment} \label{sec:spec:progress}

A major goal for most blockchains, including Tailstorm, is for the blockchain itself to grow at a constant rate so as to maintain constant transactional throughput.
However, in any deployed blockchain, the puzzle solving rate $\lambda$ changes over time because nodes may come and go or may add or remove mining hardware.
The changes in solving rate lead to changes of the growth rate.
Adjusting for the fluctuations requires feedback from the consensus to the PoW layer.
Typically, blockchains adjust the puzzle solving \emph{difficulty} depending on the observed chain growth using a dynamic \emph{difficulty adjustment algorithm} (DAA).

There exists a rich body of prior work concerning DAA design and analysis~\cite{bissias2020RadiumImproving, fullmer2018AnalysisDifficulty, harding2020RealtimeBlock, hovland2017NonlinearFeedback, kraft2016DifficultyControl}, but a deep investigation of ideal DAAs for the Tailstorm protocol is beyond the scope this paper. We note, however, that existing DAAs for Bitcoin can be adapted to Tailstorm by counting the number of subblocks where Bitcoin DAAs use the length of the blockchain.
For any Tailstorm block $b$ (summary or subblock), we define
\begin{align}
  \texttt{progress}(b) = k \cdot \texttt{height}(b) + \texttt{depth}(b) \,,
\end{align}
which counts the number of PoWs included in the chain.
Any Tailstorm DAA should adjust the puzzle solving difficulty such that \texttt{progress} grows at a constant rate.

\subsection{Protocol Variant With Constant Rewards}\label{sec:constant}

Tailstorm discounts rewards proportional to the depth of the subblock tree.
We see the discounting mechanism as a core contribution of this paper.
To isolate the effect of discounting, we introduce a protocol variant without discounting, which we call \tsconst{}.
While Tailstorm pays out \emph{at most} $c$ units of reward per subblock, the \tsconst{} protocol pays out \emph{exactly} $c$ units of reward per subblock.
In all other aspects, \tsconst{} is identical to Tailstorm.

Note, that the \tsconst{} protocol does not use the subblock tree structure, neither for consensus nor for incentives.
In that regard, \tsconst{} resembles the parallel PoW protocol $\mathcal B_k$~\cite{keller2022ParallelProofofwork} where all subblocks refer to a summary, never another subblock.
The only difference with $\mathcal{B}_k$ is that \tsconst{} does not implement leader election:
while $\mathcal B_k$ restricts the creation of the next summary to the miner of the subblock with the smallest hash,
any \tsconst{} node may (re-)create valid summaries locally.

\section{Fairness Under Protocol Compliance}
\label{sec:fairness}

A fair PoW protocol rewards miners in proportion to the amount of work they do.
Disproportionate or inconsistent allocation discriminates against weak miners and encourages the formation of pools and centralization.
In this section, we explore the causes of unfairness in Tailstorm under the assumption that all miners are honest.
We measure work in number of hashes evaluated.
The system is fair if rewards are proportional to the miners' hash rates.

The root cause of unfairness is the inherent asymmetry in reward loss when multiple PoW solutions are discovered in short order.
In Tailstorm, if more than~$k$ subblocks are produced, all having the same height, then all but~$k$ of them will be discarded.
The discarded blocks are commonly called \emph{orphans} and receive no rewards.
Typically, miners do not orphan their own blocks.
Since miners with a high hash rate are more often able to choose which blocks will be orphaned,
miners with a relatively low hash rate lose a disproportionate amount of rewards.
In Bitcoin, this effect is amplified because orphaning can occur for each block (set $k=1$ in the above argument).

\subsection{Analytical Orphan Rate Analysis}
\label{sec:orphans}

In this section, we develop an analytical model for the orphan rate and use it to compare the fairness of Bitcoin and Tailstorm. Let $B$ be a random variable denoting the number of subblocks orphaned during the production of a summary block in Tailstorm or the number of orphans per block in Bitcoin. The orphan rate is given by $\rho = E[B]/k$, which represents the expected number of PoW solutions orphaned for every PoW solution confirmed. In the remainder of this section, we will derive a bound on $\rho$.

Following Rizun's orphan rate analysis for Bitcoin~\cite{rizun2016SubchainsTechnique}, we model block propagation delays as
\begin{equation}
\tau(\tau_0, z, Q) = \tau_0 + z Q,
\end{equation}
where $\tau_0$ represents network latency (seconds), $z$ represents bandwidth (bytes per second), and $Q$ represents block size (bytes).
In order to adapt this expression for Tailstorm, we assume that the transactions included in a summary block are spread evenly across the $k$ subblocks, so the subblock size is $Q/k$. The expected summary block interval is $T$ and the subblock mining rate is $\lambda = k/T$.

\begin{restatable}{theorem}{thmOrphanRate}\label{thm:orphan_rate}
For network parameters $\tau_0$, $z$, and $Q$, summary block interval $T$, and subblock count $k$, Tailstorm's expected orphan rate~$\rho$ is bounded from above by:
\[
\rho(\tau_0, z, Q, T) \leq \frac{\tau_0}{T} + \frac{zQ}{kT}.
\]
\end{restatable}

\newcommand{\proofOrphanRate}{
\begin{proof}
Assume there are $k + X$ subblocks, all pointing to the same summary block.
Then, $k$ of these subblocks will be included in the next summary and $X$ will be orphaned.
Recall from Algorithm~\ref{alg:ts:node} that honest miners proceed to the next summary as soon as they learn of $k$ subblocks.
So, if $X$ subblocks are orphaned, they must have been mined before the $k$th subblock has propagated to all nodes.
Additionally, subblocks $k$ through $k+X$ must originate from different miners, since no miner orphans its own subblock.

Let $Y_t$ be a random variable representing the number of subblocks generated during an arbitrary time interval $t > 0$ under the assumption, introduced in Section~\ref{sec:model}, that mining intervals are exponentially distributed with rate $\lambda = k/T$. A standard result from the theory of Poisson processes shows that $Y_T \sim \texttt{Poisson}(k)$~\cite[ch.\,5]{Ross:2014}.
The same theory shows that $Y_{t} \sim \texttt{Poisson}(t \frac{k}{T})$ for arbitrary $t >0$.
According to our network model, the $k$th subblock (as well as the $k-1$ before) is fully propagated after $\tau (\tau_0 , z, Q/k)$ seconds. Thus, $X$ is equal to $Y_{\tau(\tau_0, z, Q/k)}$ in distribution.
$E[X]$~bounds the expected number of subblocks orphaned per summary block.
The orphan rate, that is subblocks orphaned per subblocks included, is bounded by
\begin{equation}
  \rho(\tau_0, z, Q, T) \leq \frac{E[X]}{k} = \frac{\tau(\tau_0 , z, Q/k) \frac{k}{T}}{k} = \frac{\tau_0}{T} + \frac{zQ}{kT}.
\end{equation}
\end{proof}
}

\proofOrphanRate{}

\begin{table}
\begin{center}
\begin{tabular}{cccccc}
\toprule
               & Bitcoin            & \multicolumn{3}{c}{Tailstorm} \\
                 \cmidrule(lr){2-2}   \cmidrule(l){3-5}
block interval & $k=1$ & $k=5$ & $k=10$ & $k=15$ \\
\midrule
$T = \phantom{0}75 $ seconds &           10.08\,\% & 7.35\,\% & 7.01\,\% & 6.89\,\% \\
$T =           150 $ seconds & \phantom{0}5.04\,\% & 3.67\,\% & 3.50\,\% & 3.45\,\% \\
$T =           300 $ seconds & \phantom{0}2.52\,\% & 1.84\,\% & 1.75\,\% & 1.72\,\% \\
$T =           600 $ seconds & \phantom{0}1.26\,\% & 0.92\,\% & 0.88\,\% & 0.86\,\% \\
\bottomrule
\end{tabular}
\end{center}
\caption{
\label{table:orphan}
Upper bounds on orphan rate obtained from Theorem~\ref{thm:orphan_rate} assuming various combinations of $k$ and expected block (Bitcoin) or summary block (Tailstorm) intervals $T$.}
\end{table}

Table~\ref{table:orphan} shows how the upper bound varies depending on the expected summary block interval~$T$ and the subblock count $k$.
The orphan rate for Bitcoin~\cite{rizun2016SubchainsTechnique} appears in the column for $k=1$.
In this analysis we assume that $Q = 32$\,MB, $z = 100$\,MB/s, and $\tau_0 = 5$\,s.
Generally, we can see that the orphan rate decreases with both expected summary block interval~$T$ and~$k$.
This suggests that, when summary block intervals are constant across protocols, Tailstorm with high~$k$ increases fairness over Bitcoin and Tailstorm with lower~$k$.

\subsection{Measuring Fairness in Simulation} \label{sec:fairness:sim}

Section~\ref{sec:orphans} provides an analytical bound for Tailstorm's orphan rate.
This bound is only tight if all subblocks originate from different miners, which is unlikely to happen in practice.
To obtain a more realistic analysis, albeit in a more specific setting, we now consider strong miners who are likely to produce multiple blocks within a single propagation delay.
Since the mathematical analysis for this scenario is complex, we rely on simulation instead.
Additionally, instead of using the orphan rate metric, itself only a proxy for fairness, we directly measure a miner's deviation from their fair share of rewards.

We implement the virtual environment as described in Section~\ref{sec:model}, Algorithm~\ref{alg:env} and measure how the choice of protocol---Bitcoin and Tailstorm under various parameterizations---affects rewards.
The network is configured with one \emph{weak} and one \emph{strong} miner operating with 1\,\% and 99\,\% of the total hash rate, respectively.
All messages are delayed for 6 seconds.

We simulate one million PoW solutions per configuration.
We split the simulation into \emph{independent observations}, each representing one day of protocol execution.
Specifically, with $T$ denoting the expected summary block interval, we terminate individual executions when $\lfloor 24 \times 3600 \times k / T \rfloor$ PoWs are mined.
For each execution, we identify the longest chain of blocks and calculate the accumulated rewards for both miners.
Since each observation covers one day of mining, the variance represents daily volatility of rewards.

\begin{figure}
  \includegraphics[width=\linewidth]{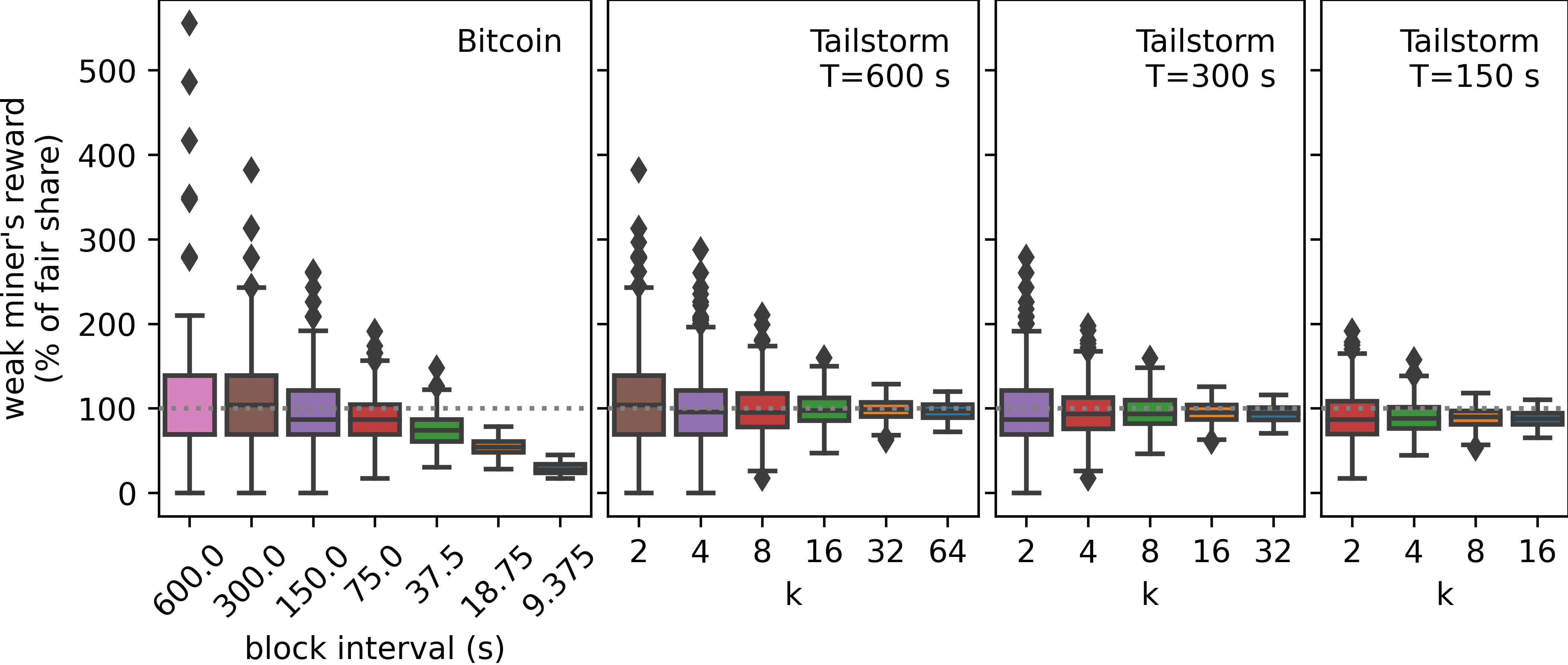}
  \caption{
    Observed fairness for different configurations of Bitcoin (leftmost facet) and Tailstorm (right facets) with expected summary interval $T$ ranging from 600 to 150 seconds. Colors represent expected subblock mining intervals.
  }
  \label{fig:inequality}
\end{figure}

A miner's fair share of reward equals its relative hash rate.
To measure fairness, we compare actual rewards to the fair share.
Figure~\ref{fig:inequality} shows the weak miner's relative reward as a percentage of its fair share.
The leftmost facet shows the Bitcoin protocol for expected block intervals ranging from roughly 9 seconds to 600 seconds.
The remaining three facets show Tailstorm with \emph{summary} block intervals $T$ of 600, 300, and 150 seconds for $k$ ranging from 2 to 64.
Like colors represent such configurations where \emph{subblocks} are generated at the same frequency~$\lambda = k/T$.
We omit the configurations where the expected subblock interval $T/k$ is lower than the network propagation delay of 6 seconds.

The Figure illustrates the tradeoff between reward fairness and volatility in Bitcoin and Tailstorm. In the leftmost facet, it shows how increasing the expected block interval improves fairness but also increases volatility in Bitcoin.
In contrast, the right facets show how Tailstorm can be configured to  independently control both reward fairness (by adjusting the summary block interval $T$) and volatility (by adjusting $k$).
In particular, choosing higher $k$ leads to more frequent rewards and lower volatility, while longer summary block intervals tend to increase fairness.
The latter observation aligns with the results in Table~\ref{table:orphan}, which also shows that fairness tends to increase with the summary block interval.

\section{Attack Evaluation} \label{sec:attacks}

We now extend the virtual environment defined by Algorithm~\ref{alg:env} in Section~\ref{sec:model} to model adversarial behavior.
We then specify and evaluate several attack strategies against Tailstorm.

To account for the possibility of collusion among multiple dishonest parties, we pessimistically assume that they indeed collude.
Therefore, we model a single but strong node deviating from the protocol.
Throughout the remaining sections we refer to the single dishonest node as the \emph{attacker} and to the honest nodes as \emph{defenders}.
The attacker \emph{observes} the virtual environment's state and reacts to any updates accordingly.
We describe attacks as \emph{policies}, which determine the \emph{action} to take based on the observed state.

\subsection{Network} \label{sec:attacks:network}

We deploy a virtual environment comprising $n$ nodes.
The first node represents the attacker and is assigned fraction $\alpha$ of the total hash rate.
The remaining hash rate is distributed evenly among the $n - 1$ defenders.
Henceforth, we refer to $\alpha$ as the \emph{attacker's strength}.

In Bitcoin, blocks form a linear chain.
If there are two blocks with the same height, only one will be included in the blockchain long term.
Since honest Bitcoin nodes prefer and mine on the block first received, nodes with better connectivity obtain an advantage.
The same \emph{block race} situation can arise in Tailstorm.
Here, there can be only one \emph{summary} at each height and the nodes' preference of summaries (Alg.\,\ref{alg:ts:node}) likewise depends on the order they are received.
Attackers able to send blocks more quickly might be able to manipulate the defenders' preference to suit their needs.

We follow Eyal and Sirer~\cite{eyal2014MajorityNot} and Sapirshtein et al.~\cite{sapirshtein2016OptimalSelfish} who model the \emph{block race advantage} with a single network parameter~$\gamma$, which defines the proportion of defenders, weighted by hash rate, that will opt for the attacker's block in the case of a block race.
We pessimistically assume that the attacker receives all blocks without delay.
Defender nodes can send blocks to one another with a negligible delay of $\varepsilon$.
When the attacker sends a block to a defender, we introduce a random delay, which we draw independently from a uniform distribution on the interval $[0, \frac{n-2}{n-1}\frac{\varepsilon}{\gamma}]$.

\begin{restatable}{theorem}{thmGammaAligns}\label{thm:gamma_aligns}
Let $\gamma \in [0, 1]$ be given. For any choice of $\varepsilon > 0$ and $n > \frac{1}{1-\gamma} + 1$, $\gamma$ is equal to the attacker's block race advantage.
\end{restatable}

\newcommand{\proofGammaAligns}{
\begin{proof}
Let $A$ be the event that a randomly chosen honest miner accepts the attacker block over honest and $A_i$ the event that a specific miner accepts the attacker block. By the law of total probability we have $P[A] =  \frac{1}{n-1} \sum_{i=1}^{n-1} P[A_i]$. Note also that if we assume w.l.o.g.~that $A_{n-1}$ was the miner of the competing block, then $P[A_{n-1}] = 0$. Since defenders communicate with each other with delay $\varepsilon$ while the attacker communicates with defenders with uniform random delay up to time $\frac{n-2}{n-1}\frac{\varepsilon}{\gamma}$, it must be the case that
\begin{align}
  P[A_i] = \varepsilon \left(\frac{n-2}{n-1}\frac{\varepsilon}{\gamma}\right)^{-1}\,,\quad\forall i < n-1\,,
\end{align}
provided that $\frac{n-2}{n-1}\frac{\varepsilon}{\gamma} > \varepsilon$. Choosing
$n > \frac{1}{1-\gamma} + 1$ satisfies this  constraint.
Equality in the constraint is not valid because it forces $P[A_i] = 1$ for all~$\gamma$.
Therefore,
\begin{align}
P[A] &= \frac{1}{n-1} \sum_{i=1}^{n-1} P[A_i] \\
     &= \frac{1}{n-1} \sum_{i=1}^{n-2} P[A_i] \\
     &= \frac{n-2}{n-1} \frac{\varepsilon}{\frac{n-2}{n-1}\frac{\varepsilon}{\gamma}} \\
     &= \gamma.
\end{align}
\end{proof}
}

\proofGammaAligns{}

Constraint $n > \frac{1}{1-\gamma} + 1$, from Theorem~\ref{thm:gamma_aligns}, is fundamental; if $n$ is smaller, then it is not possible for an honest miner to communicate a block to another honest miner faster than the attacker can. The constraint implies that $\gamma < \frac{n-2}{n-1}$.
Hence, we can model $\gamma = 1$ only in the limit that the number of miners approaches infinity, which is not possible in practice.
Other authors~\cite{eyal2014MajorityNot, sapirshtein2016OptimalSelfish} have directly considered $\gamma = 1$, but we feel that the restriction to $\gamma < 1$ is natural.
Given that the miner of the defending block in a block race will never adopt the attacker's competing block, $\gamma = 1$ implies that individual miners have zero hash rate, which is not realistic.

\subsection{Observation Space} \label{sec:attacks:obs}

The virtual environment maintains complex state.
It tracks the full history of blocks, delayed messages, and the partial views of all nodes.
Some of this information is not available to attackers in practice, other parts are not relevant.
To enable concise policies, we follow Sapirshtein et al.~\cite{sapirshtein2016OptimalSelfish} and restrict what the attacker can see.

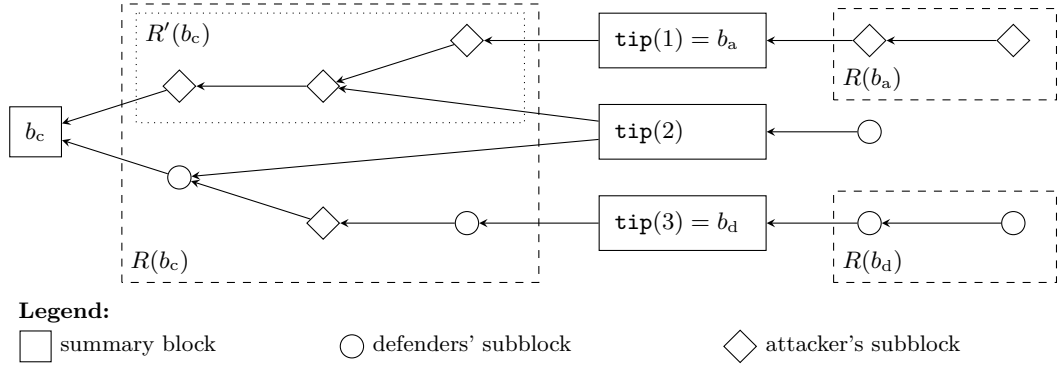
\begin{figure}
  \begin{tikzpicture}[
  >=stealth,
  x=\linewidth/7.4,
  y=4ex,
  sba/.style={draw, diamond},
  sbd/.style={draw, circle},
  sum/.style={draw, rectangle, inner sep=1.5ex},
  ]
  \small
  \node[sum] (s0) at (0.0,3.5) {$b_\text c$};
  \node[sba] (a1) at (1.0,4.5) {};
  \node[sba] (a2) at (2.0,4.5) {};
  \node[sba] (a3) at (3.0,5.5) {};
  \node[sbd] (d1) at (1.0,2.5) {};
  \node[sba] (d2) at (2.0,1.5) {};
  \node[sbd] (d3) at (3.0,1.5) {};
  \node[sum, text width=5.5em] (s1) at (4.5,5.5) {$\texttt{tip}(1) = b_\text a$};
  \node[sum, text width=5.5em] (s2) at (4.5,3.5) {$\texttt{tip}(2)$};
  \node[sum, text width=5.5em] (s3) at (4.5,1.5) {$\texttt{tip}(3) = b_\text d$};
  \node[sba] (a4) at (5.8,5.5) {};
  \node[sba] (a5) at (6.8,5.5) {};
  \node[sbd] (d4) at (5.8,3.5) {};
  \node[sbd] (d5) at (5.8,1.5) {};
  \node[sbd] (d6) at (6.8,1.5) {};
  \draw[->] (a1) -- (s0);
  \draw[->] (a2) -- (a1);
  \draw[->] (a3) -- (a2);
  \draw[->] (s1) -- (a3);
  \draw[->] (s2) -- (a2);
  \draw[->] (s2) -- (d1);
  \draw[->] (s3) -- (d3);
  \draw[->] (d1) -- (s0);
  \draw[->] (d2) -- (d1);
  \draw[->] (d3) -- (d2);
  \draw[->] (d4) -- (s2);
  \draw[->] (d5) -- (s3);
  \draw[->] (d6) -- (d5);
  \draw[->] (a4) -- (s1);
  \draw[->] (a5) -- (a4);
  \draw[dashed] (0.6, .2) node[above right]{$R(b_\text c)$} rectangle (3.5, 6.3);
  \draw[dotted] (0.7, 6.1) node[below right]{$R'(b_\text c)$} rectangle (3.4, 3.7);
  \draw[dashed] (5.55, 0.2) node[above right]{$R(b_\text d)$} rectangle (7.1, 2.2);
  \draw[dashed] (5.55, 4.2) node[above right]{$R(b_\text a)$} rectangle (7.1, 6.2);


  \begin{scope}[shift={(0,-1.2)}]
    \node[] at (.2, 0.75) {\footnotesize\textbf{Legend:}};
    \node[sum, label=right:{\footnotesize\strut summary block}] at (0,0) {};
    \node[sbd, label=right:{\footnotesize\strut defenders' subblock}] at (2.2,0) {};
    \node[sba, label=right:{\footnotesize\strut attacker's subblock}] at (4.9,0) {};
  \end{scope}
\end{tikzpicture}
  \caption{%
    Notation in Section~\ref{sec:attacks:obs} Observation Space.
    This example uses three nodes and $k=3$.%
  }
  \label{fig:obsspace}
\end{figure}

Figure~\ref{fig:obsspace} illustrates our notation.
At any given time, let $b_\text{a} = \texttt{tip}(1)$ denote the attackers preferred summary block, and let $b_\text{d}$ denote the best block (according to \texttt{Preference} in Alg.\,\ref{alg:ts:node})
 among the preferred blocks of the defenders: $\texttt{tip}(2), \ldots, \texttt{tip}(n)$.
We use $b_\text{c}$ to refer to the best summary block among the common ancestors of $b_\text{a}$ and $b_\text{d}$.
In other words, $b_\text{c}$ is the latest block that \emph{all} nodes agree upon.
For any summary block $b$, we use $R(b)$ to identify the subblocks which are both a descendant of $b$ and have the same height as $b$. Note that Tailstorm's chain structure (Alg.\,\ref{alg:ts:valid}) implies that $R(b)$ and $b$ together span a tree.
Section~\ref{sec:spec} defines the \emph{depth} of a subblock to be its depth within this tree.
We define~$R'(b)$ as the largest subset of $R(b)$ such that all subblocks are mined by the attacker and $R'(b) \cup \{b\}$ is still connected.
In the following, we use $|S|$ to refer to the cardinality of a set $S$ and $\texttt{depth}[S]$ to refer to the maximum depth among all blocks in~$S$.
The attacker observes

\bgroup
  \newcommand{\obsvar}[3]{%
    \item[$#1$:] the #2, $#3$
  }
  \begin{description}
      \obsvar{h_\text{a} }{attacker's height advantage}{\texttt{height}(b_\text{a}) - \texttt{height}(b_\text{c})},
      \obsvar{h_\text{d} }{defenders' height advantage}{\texttt{height}(b_\text{d}) - \texttt{height}(b_\text{c})},
      \obsvar{s_\text{a} }{attacker's inclusive subblock count}{|R(b_\text{a})|},
      \obsvar{s_\text{a}'}{attacker's exclusive subblock count}{|R'(b_\text{a})|},
      \obsvar{s_\text{d} }{defenders' subblock count}{|R(b_\text{d})|},
      \obsvar{d_\text{a} }{attacker's inclusive depth}{\texttt{depth}[R(b_\text{a})]},
      \obsvar{d_\text{a}'}{attacker's exclusive depth}{\texttt{depth}[R'(b_\text{a})]}, and
      \obsvar{d_\text{d} }{defenders' depth}{\texttt{depth}[R(b_\text{d})]}.
  \end{description}

\egroup 

\def\obsvars{
h_\text{a} ,
h_\text{d} ,
s_\text{a} ,
s_\text{a}',
s_\text{d} ,
d_\text{a} ,
d_\text{a}',
d_\text{d} }

\noindent The example shown Figure~\ref{fig:obsspace} implies
$
  (\obsvars) = (1,1,2,2,2,2,2,2)\,.
$

\subsection{Action Space} \label{sec:attacks:actions}

\begin{algorithm}[t]
  \caption{Tailstorm: Attacker}
  \label{alg:ts:attacker}

  \BlankLine
  \Fn(\tcc*[f]{$t$: currently preferred block // $b$: new block}){$\texttt{Update}(\text{tip } t, \text{block } b)$}{
    $\emph{pref}, \emph{share}, \emph{append} \gets t, [], []$%
    \tcc*[r]{new preferred tip, blocks to broadcast, templates to append}
    \lIf{$b$ is own summary block}{$\emph{pref} \gets b$}
    $W \gets$ set of withheld blocks (= blocks not shared before)\;
    $\emph{withhold}, \emph{extend} \gets \mathcal P(\obsvars)$%
    \tcc*[r]{$\mathcal P$: attack policy}
    \Switch{\emph{withhold}}{
      \lCase{\texttt{Adopt}}{
        $\emph{pref} \gets b_\text{d}$%
        \tcc*[f]{$b_\text d$: defender's block, see Sec.\,\ref{sec:attacks:obs}}
      }
      \lCase{\texttt{Match}}{%
        $\emph{share} \gets \left\{ x \in W \mid \texttt{progress}(x) \leq \texttt{progress}(b_\text{d}) \right\}$
      }

      \lCase{\texttt{Override}}{%
        $\emph{share} \gets \left\{ x \in W \mid \texttt{progress}(x) \leq \texttt{progress}(b_\text{d}) + 1 \right\}$
      }
      \lCase(){\texttt{Wait}}{nothing}
    }
    \Switch{\emph{extend}}{
      \Case{\texttt{Inclusive}}{
        \If(\tcc*[f]{$R(b)$: full subblock tree confirming $b$, see Sec.\,\ref{sec:attacks:obs}}){$|R(\emph{pref\,})| \geq k$}{
          $\emph{append} \gets$ summary template using $R(\emph{pref\,})$\;
        }
      }
      \Case{\texttt{Exlusive}}{
        \If(\tcc*[f]{$R'(b)$: partial subblock tree confirming $b$, see Sec.\,\ref{sec:attacks:obs}}){$|R'(\emph{pref\,})| \geq k$}{
          $\emph{append} \gets$ summary template using $R'(\emph{pref\,})$\;
        }
      }
    }
    \Return \emph{pref}, \emph{share}, \emph{append}\;
  }
\end{algorithm}

The virtual environment (Alg.\,\ref{alg:env}) calls a node's \texttt{Update} function whenever this node learns about a new block: after it was mined locally with PoW, appended locally without PoW, or received from the network.
We implement the attacker's \emph{dishonest} \texttt{Update} function in Algorithm~\ref{alg:ts:attacker}.
We allow for generic attacks by letting the attacker choose from a set of potentially dishonest actions.
We assume that there is a policy $\mathcal P$ which maps observations (see Sect.\,\ref{sec:attacks:obs}) to action tuples of the form (\emph{withhold}, \emph{extend}).
The \emph{withhold} action type controls the preferred tip of the chain and the withholding of blocks.
We hereby follow closely the actions used by Sapirshtein et al.\,\cite{sapirshtein2016OptimalSelfish} for selfish mining against Bitcoin~\cite{eyal2014MajorityNot}.

\begin{description}
  \item[\texttt{Wait}] Continue mining on $b_\text a$ and withhold new blocks.
  \item[\texttt{Match}] Release just enough blocks to induce a block race between $b_\text d$ and an attacker block.
  \item[\texttt{Override}] Release just enough blocks to make the defenders discard their block $b_\text d$.
  \item[\texttt{Adopt}] Abort attack, prefer defenders' block $b_\text d$, and discard $b_\text a$.
\end{description}

Recall from Section~\ref{sec:model} that ``blocks'' refers to vertices in the block DAG.
This makes the \emph{withhold} action protocol-agnostic.
For Tailstorm, blocks can be subblocks or summaries.
Note that while \texttt{Adopt} and \texttt{Wait} are always feasible, \texttt{Match} and \texttt{Override} are not.
If the attacker's chain is too short, the defenders will not consider adopting it.
In such cases, \texttt{Match} and \texttt{Override}, as implemented in Algorithm~\ref{alg:ts:attacker}, fall back to releasing all withheld blocks.

The second action type, \emph{extend}, is specific to Tailstorm.
It controls how the attacker assembles new summary blocks, more specifically, which subblocks it considers for selection with Algorithm~\ref{alg:subblock_selection}.

\begin{description}
  \item[\texttt{Inclusive}] Use all available subblocks to create new summaries.
  \item[\texttt{Exclusive}] Use only subblocks which were mined by the attacker.
\end{description}

Note that the attacker never delays the next summary block longer than necessary.
As soon as there are enough subblocks for an inclusive or exclusive summary (according to the chosen action), this summary will be created.

\subsection{Reference Policies}\label{sec:attacks:policies}

We evaluate the following policies.
In each, the attacker assembles \texttt{Inclusive} summaries.

\begin{description}
  \item[Honest]
    Emulate Algorithm~\ref{alg:ts:node}: adopt the longest chain and release all blocks:\par
    \texttt{Adopt} if $h_\text{d} > h_\text{a}$. Otherwise \texttt{Override}.

  \item[Get Ahead]
    Withhold own subblocks, release own summaries:\par
    \texttt{Adopt} if $h_\text{d} > h_\text{a}$.
    \texttt{Override} if $h_\text{d} < h_\text{a}$.
    Otherwise \texttt{Wait}.

  \item[Minor Delay]
    Withhold own subblocks, override defender summaries as they come out:\par
    \texttt{Adopt} if $h_\text{d} > h_\text{a}$.
    \texttt{Wait} if $h_\text{d} = h_\text{c}$.
    Otherwise \texttt{Override}.

\end{description}

To evaluate the policies above, we reuse the simulator from Section~\ref{sec:fairness:sim}, configuring the network as described in Sections~\ref{sec:attacks:network} to~\ref{sec:attacks:actions}.
We set block race advantage $\gamma \in \{5, 50, 95\}\,\%$ and allow the attacker strength $\alpha$ to range from 20\,\% to 45\,\%.
We run one hundred simulations per configuration, and stop each as soon as the DAG contains 2048 blocks.
At the end of each simulation we select the longest chain of blocks and calculate its rewards.
We configure Tailstorm's discount reward scheme, as defined in Section~\ref{sec:spec:chain}, with $c = 1$ such that at most one unit of reward is minted per unit of chain progress.

Some policies cause more orphans than others.
Assuming effective difficulty adjustment according to Section~\ref{sec:spec:progress}, the amount of simulated time depends on the attacker's behaviour.
The attacker's mining cost is proportional to the time spent on the attack.
To facilitate comparison across policies, we normalize rewards with respect to simulated time.
Formally, we define the \emph{normalized reward} as the attacker's reward up to block~$b$ divided by $\texttt{progress}(b)$.
Note that relative reward, a metric commonly used for Bitcoin~\cite{barzur2020EfficientMDP, eyal2014MajorityNot, sapirshtein2016OptimalSelfish, zhang2019LayCommon}, is not sufficient to account for Tailstorm's reward discounting.

\begin{figure}
  \centering
  \includegraphics[width=\linewidth]{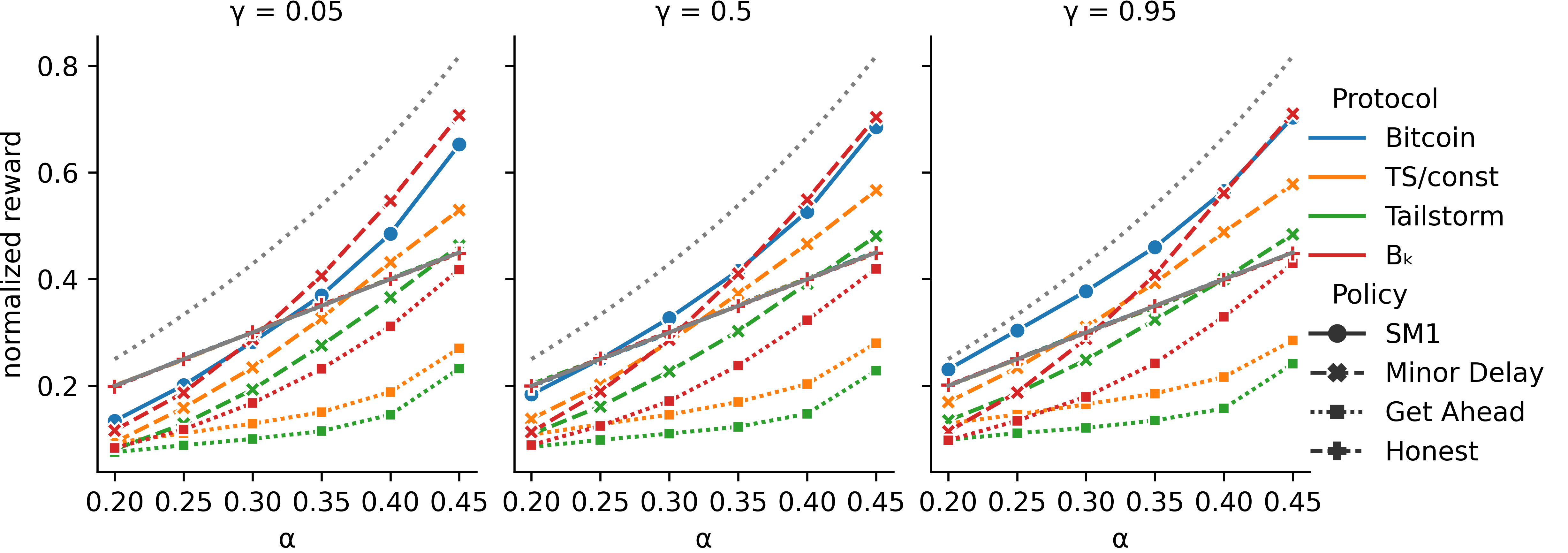}
  \caption{
    Observed normalized reward as a function of attacker strength~$\alpha$ for different protocols (color) and reference policies (line style and markers).
    The three facets represent the network assumptions $\gamma \in \{5, 50, 95\}\,\%$.
    We set protocol parameter $k=8$ where applicable.
    \tsconst{} is Tailstorm without reward discounting.
    We add a gray reference line for the fair reward $\alpha$ (overlaps with honest policy) and a gray dotted curve for the known upper bound $\alpha / (1 - \alpha)$.
  }
  \label{fig:hard_coded}
\end{figure}

Figure~\ref{fig:hard_coded} reports the average normalized reward  (across 100 simulations) on the y-axis for the varying $\alpha$ on the x-axis and with $\gamma$ varying by facet.
The green curves represent the different reference policies against Tailstorm.
We also evaluate the reference policies against the \tsconst{} protocol as defined in Section~\ref{sec:constant} (orange curves) and against $\mathcal B_k$ as defined in Appendix~\ref{apx:bk} (red curves).
In addition, we evaluate the SM1 policy against Bitcoin (blue curve) as described by Sapirshtein et al.~\cite{sapirshtein2016OptimalSelfish} and in Appendix~\ref{apx:bitcoin}.
For orientation we include their upper bound $\alpha / ( 1 - \alpha)$ as a gray dotted curve and add a solid gray line for $\alpha$, the expected reward of honest behaviour.

The figure supports multiple conclusions.
First, the curves for the \emph{Honest} policy coinciding with the line for $\alpha$ indicates that the policy indeed replicates honest behavior in all evaluated protocols.
Second, comparing the measurements for Tailstorm with the ones for \tsconst{} shows clearly that the discounting of rewards reduces efficacy of all dishonest reference policies.
Third, \emph{Minor Delay} is generally the most effective reference policy.
Forth, \emph{Minor Delay} produces less reward in Tailstorm than SM1 does in Bitcoin.
Fifth, for low~$\gamma$, \emph{Minor Delay} is more profitable against $\mathcal B_k$ than SM1 is against Bitcoin.

Note however, that $\emph{Minor Delay}$ might not be the best strategy against Tailstorm, just like
SM1 is not always optimal for Bitcoin~\cite{sapirshtein2016OptimalSelfish}.
This motivates to search for optimal policies.

\section{Attack Search}\label{sec:reinforcement_learning}

Previous research has identified optimal attacks against Bitcoin and Ethereum through the use of Markov Decision Processes (MDP) and exhaustive search~\cite{barzur2020EfficientMDP, gervais2016SecurityPerformance, sapirshtein2016OptimalSelfish}. However, for more complex protocols, the state space of MDPs can become prohibitively large, and exhaustive search becomes impractical~\cite{hou2021SquirRLAutomating, keller2022ParallelProofofwork, zhang2019LayCommon}. As a result, many authors in the past have resorted to evaluating hard-coded attacks instead of searching for optimal policies~\cite{bissias2020BobtailImproved, feng2019SelfishMining, keller2022ParallelProofofwork, kiffer2018BetterMethod, marmolejocossio2019CompetingSemi}.

We adopt the approach introduced by Hou et al.~\cite{hou2021SquirRLAutomating} and employ reinforcement learning (RL) to search for attacks.
The observation and action spaces described in Section~\ref{sec:attacks} readily define a partially observable MDP.
To search for optimal attacks, we replace the hard-coded policy $P$ with an RL agent that learns to select actions based on past observations.

We utilize off-the-shelf RL tooling by first exposing our simulator and attack space as an OpenAI Gym~\cite{brockman2016OpenAIGym}. Next, we deploy Proximal Policy Optimization (PPO) \cite{schulman2017ProximalPolicy} as our agent. To support reproduction and future research, we release the Gym as Python package~\anoncite{cpr:pypi} and include all training scripts in our open source repository~\anoncite{cpr:repo}.

The modularity of our simulator allows us to apply the same RL pipeline to different protocols.
We insert the protocol specifications along with their associated observation and action spaces, while reusing the virtual environment and network assumptions across protocols.
For Bitcoin, we adopt the attack space defined by Sapirshtein et al.~\cite{sapirshtein2016OptimalSelfish}.
Details about the attack spaces for Bitcoin and $\mathcal B_k$ are provided in Appendix~\ref{apx:protocols}.

We use Bitcoin as a reference protocol to measure the completeness of our approach.
Previous research by Sapirshtein et al.~\cite{sapirshtein2016OptimalSelfish} has yielded the optimal policy.
As we will demonstrate, our search mechanism reproduces these results closely.

We search for optimal policies for all possible combinations of attacker strength $\alpha \in \{20, 25, 30, \dots 45\}$\,\%, and block race advantage $\gamma \in \{ 5, 50, 95\}$\,\%.
We evaluate four protocols: Bitcoin, $\mathcal B_k$, Tailstorm, and \tsconst{} with $k=8$ where applicable.
In total, this amounts to $7 \cdot 3 \cdot 4 = 84$ different learning problems.

For each learning problem, we conduct multiple training runs with varying hyperparameters.
For the objective function we choose \emph{normalized reward}, as evaluated for the reference policies in Figure~\ref{fig:hard_coded}.
From the training we obtain $1292$ policies.

We select the best trained policy for each problem by simulating 100 independent protocol executions for each and proceeding as follows.
As in Section~\ref{sec:attacks:policies}, we stop individual executions as soon as there are 2048 blocks.
We then select the longest chain of blocks and observe the attacker's normalized reward.
This reward is averaged over the 100 observations per policy to determine the policy with the highest average reward.
For reference, we apply the same filtering method to select the best policy among the reference policies from Section~\ref{sec:attacks:policies}.

\begin{figure*}
  \includegraphics[width=\linewidth]{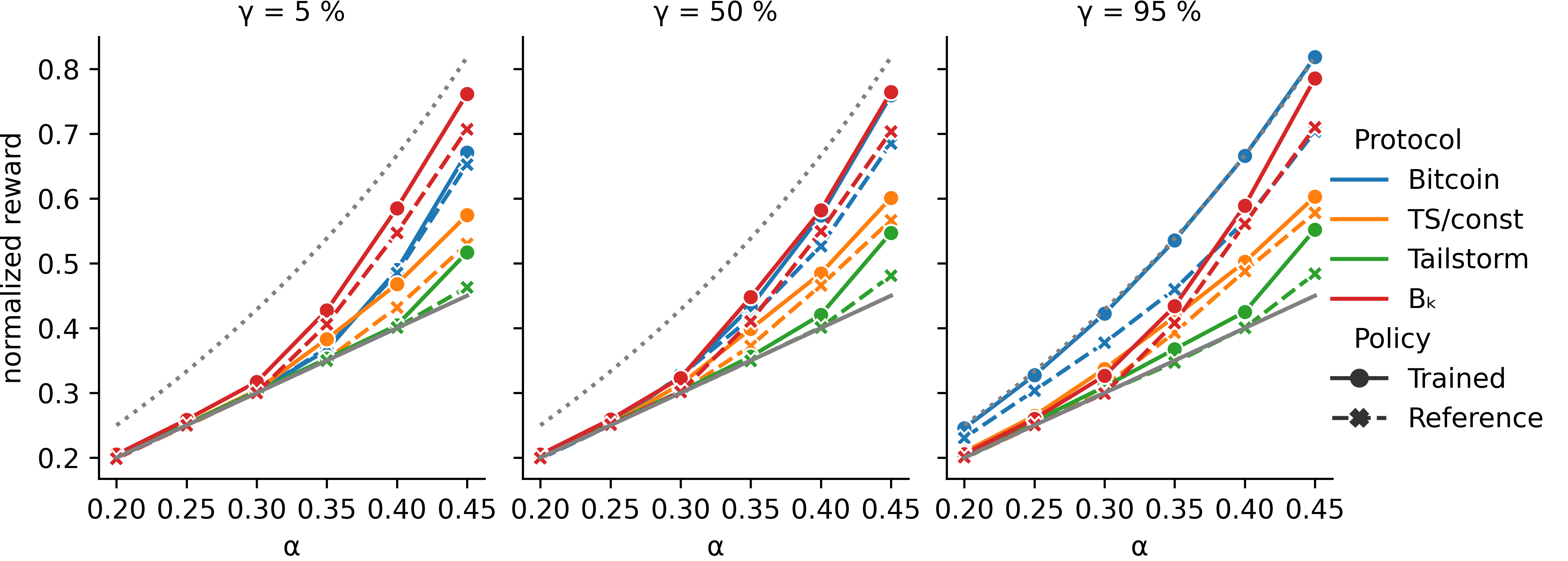}
  \caption{
    Observed normalized reward of the \emph{best known policy} with the axes set up like in Figure~\ref{fig:hard_coded}.
    Solid curves with $\bullet$-markers represent trained reinforcement learning models.
    Dashed curves with $\times$-markers represent the reference policies defined in Section~\ref{sec:attacks:policies}.
  }
  \label{fig:rl}
\end{figure*}

Figure~\ref{fig:rl} shows the performance of the selected policies, with $\gamma$ varying by facet, $\alpha$ on the x-axis, and average normalized reward on the y-axis.
The solid colored curves represent the best trained policies, while the dashed colored curves represent the best reference policies.
As in Figure~\ref{fig:hard_coded}, we include gray reference curves for Bitcoin's known upper bound  $\alpha / (1 - \alpha)$ (dotted) and fair share $\alpha$ (solid).
Note that for $\gamma = 50\,\%$, Nakamato and $\mathcal B_k$ visually overlap.

In comparing the performance of different protocols, we say that  protocol $A$ performs \emph{better} (or \emph{worse}) than~$B$ in the event that~$A$ returns fewer (alternatively \emph{more}) rewards to the attacker than does~$B$. Figure~\ref{fig:rl} supports the following conclusions.
First, $\mathcal B_k$ performs worse than Bitcoin for low~$\gamma$, but better than Bitcoin for high $\gamma$.
The break-even point seems to be at $\gamma = 50\,\%$.
Second, Tailstorm consistently outperforms Bitcoin and $\mathcal B_k$, even with constant rewards (\tsconst{}).
Moreover, discounting of rewards consistently makes Tailstorm less susceptible to selfish mining and similar incentive layer attacks.
Third, the learned policies consistently match or outperform the hard-coded reference policies.
This finding is important because anything else would indicate deficiencies in the training.
Forth, the learned policies are close to optimal for Bitcoin.
This can be seen in two ways.
Firstly, Sapirshtein et al. showed that the SM1 policy is close to optimal for $\gamma = 0$~\cite[Fig.\,1a]{sapirshtein2016OptimalSelfish}.
Our figure for $\gamma = 5\%$ shows that the learned policy matches SM1 (the dashed blue curve with $\times$-markers).
Secondly, the authors showed that the optimal policy reaches the upper bound $\alpha / (1 - \alpha)$ for $\gamma = 1$~\cite[Fig.\,1c]{sapirshtein2016OptimalSelfish}.
Our figure for $\gamma = 95\%$ shows that the learned policy matches this bound as well.

Until now, we have evaluated the efficacy of policies in absolute terms.
We now take a different perspective and ask: how strong must an attacker be for dishonest behaviour to pay off?
Recall that honest behaviour implies an expected normalized reward of~$\alpha$ (solid gray line in Figures~\ref{fig:hard_coded} and~\ref{fig:rl}).
To answer the question, we calculate break-even points, which represent the minimum relative hash rate $\alpha$, such that following the policy produces more than $\alpha$ of normalized reward.
We start from the optimal policies presented in Figure~\ref{fig:rl} and select only those policies that feature dishonest behavior.
Recall that each of the remaining policies was trained on a fixed relative hash rate $\alpha$.
For each protocol and choice of  $\gamma$, we select the dishonest policy trained for the lowest~$\alpha$.
We then evaluate this policy against alternative $\alpha$ values ranging from 5\,\% to 50\,\%.
Our simulator provides noisy observations and the reward distribution varies stochastically.
Hence, we use Bayesian optimization to minimize the difference between the observed reward and $\alpha$, i.\,e., to get as close to the break-even point as possible.
Table~\ref{tab:breakeven} reports the results.
Observe that for the trained policies, the break-even points are consistently either lower or close to the break-even points of the hard-coded reference policies.
$\mathcal B_k$, \tsconst{}, and Tailstorm are also less sensitive to changes in $\gamma$ than Bitcoin.
Tailstorm has the highest break-even points among all protocols, indicating that it is most resilient to incentive layer attacks.

\begin{table}
  \caption{
    Break-even points: minimal $\alpha$ (in\,\%) where dishonest behaviour (Figure~\ref{fig:rl}) pays off.
  }
  \label{tab:breakeven}
  \bgroup
\newcolumntype{Y}{>{\centering\arraybackslash}X}
\begin{tabularx}{\linewidth}{lYYYYYY}
\toprule
 & \multicolumn{3}{c}{Trained Policy} & \multicolumn{3}{c}{Reference Policy} \\
\cmidrule(lr){2-4}\cmidrule(l){5-7}
$\gamma$ & 5\,\% & 50\,\% & 95\,\% & 5\,\% & 50\,\% & 95\,\% \\
\midrule
Bitcoin & 33.6 & 26.5& {$\leq 5$} & 32.7 & 24.5& {$\leq 5$} \\
\tsconst{} & 28.9 & 28.6 & 21.7 & 37.9 & 32.6 & 28.4 \\
Tailstorm & 41.6 & 35.8 & 35.8 & 43.7 & 41.0 & 39.3 \\
$\mathcal B_k$ & 22.2 & 27.8 & 25.3 & 31.3 & 31.1 & 30.7 \\
\bottomrule
\end{tabularx}
\egroup

\end{table}

\section{Tailstorm Cryptocurrency}
\label{sec:implementation}

So far, we have focused on Tailstorm's consensus and incentive mechanisms.
Throughout our analysis, we have assumed a cryptocurrency on the application layer that facilitates the creation and distribution of rewards to participants of the consensus layer.
In this section, we describe and discuss the Tailstorm cryptocurrency and our prototype implementation.

\subsection{Transaction Handling} \label{sec:transaction}

Tailstorm implements the unspent transaction output (UTXO) model as it is used in Bitcoin~\cite{nakamoto2008BitcoinPeertopeer}:
Each UTXO represents a designated amount of cryptocurrency.
Transactions consume and create UTXOs, but they never create more cryptocurrency than they consume.
Consumed UTXOs cannot be consumed again.
Ownership and transfer of value follows from restricting the consummation of UTXOs to the holders of specific cryptographic keys.

Tailstorm uses the same public key cryptography as Bitcoin, and it also supports Bitcoin's UTXO scripting facility.
The dissemination and processing of transactions follows Bitcoin's approach as well.
However, as we describe next, Tailstorm deviates from Bitcoin in how transactions are recorded in the blockchain.

\bgroup
\newcommand{\tx}[1]{\ensuremath{\texttt{tx}_{\text{#1}}}}
\newcommand{\blk}[1]{\ensuremath{\texttt{sub}_{\text{#1}}}}
\newcommand{\idx}[1]{\ensuremath{\texttt{idx}_{\text{#1}}}}
\newcommand{\smy}[1]{\ensuremath{\texttt{sum}_{\text{#1}}}}
\newcommand{\dst}[1]{\ensuremath{\texttt{dst}_{\text{#1}}}}

Taking inspiration from Storm~\cite{awemany2019StormWhitepaper}, each subblock contains a list of transactions.
Let \tx{a} be a transaction listed at position \idx{a} in subblock \blk{a} summarized in summary block \smy{a} and let
\tx{b}, \idx{b}, \blk{b}, and \smy{b} be defined similarly.
Assume \tx{a} and \tx{b} are incompatible, e.\,g., because they spend the same UTXO twice.
If \tx{a} and \tx{b} are listed in the same subblock, i.\,e., $\blk a = \blk b$ and $\idx a \neq \idx b$, then this subblock is marked invalid.
We proceed similarly if \tx a  and \tx b are summarized by different summaries:
due to Tailstorm's chain structure (see Alg.\,\ref{alg:ts:valid}) \smy a and \smy b must differ in height and we mark the higher one as invalid.
Honest nodes ignore invalid blocks and all their descendants at the consensus layer.
Hence, under the above circumstances, incompatible transactions are not persisted in the blockchain, as is the case in Bitcoin.

However, special attention is required for incompatible transactions in different subblocks summarized by the same summary block, i.\,e., $\blk a \neq \blk b$ and $\smy a = \smy b$.
A critical assumption of our analyses as well as parallel PoW in general~\cite{keller2022ParallelProofofwork} is that subblocks confirming the same summary are compatible and thus can be mined independently.
We thus cannot mark the summary invalid in this case and the incompatible transactions are persisted in the blockchain.
To resolve this conflict, Tailstorm executes only one of the transactions at the application layer and ignores the other according the following rules.
Let \dst a denote the distance of \blk a to \smy a in the block DAG, and let \dst b be defined similarly.
If $\dst a < \dst b$, then \tx a is ignored.
Ties are broken by the PoW hash function, i.\,e., if $\dst a = \dst b$ and $\texttt{hash}(\blk a) > \texttt{hash}(\blk b)$, then \tx a is ignored.
If \tx a is ignored then all dependent transactions (i.e. those spending the UTXOs produced by \tx a) are ignored as well.
Note that even though incompatible transactions are persisted in the blockchain, the double-spending semantics are equivalent to Bitcoin: offending transactions off the longest branch are ignored.

\subsection{Fast Confirmations}

To reap the full consistency guarantees of parallel PoW~\cite{keller2022ParallelProofofwork}, prudent cryptocurrency users should wait for one summary block confirmation before accepting their transactions as final.
For example, if their transaction is included in subblock~\blk a and first summarized in \smy a, then they should wait for another valid summary \smy b with $\texttt{height}(\smy b) = \texttt{height}(\smy a) + 1$.
Assuming a 10 minute summary block interval and large $k$, the full confirmation will likely occur in 10 to 20 minutes, depending on whether the transaction was included early in the subblock tree or later.

If time is short and the transacted value is low, e.\,g.\ if the user is selling a cup of coffee to go, they can consider waiting for a number of subblock confirmations instead.
For example, consider Tailstorm with $k = 60$ and a 10 minute summary block interval.
Subblocks will be mined  every 10 seconds in expectation.
The heuristics are similar to a fast version of Bitcoin:
If the seller waits for 6 subblock confirmations, i.\,e., for a subblock \blk b with $\texttt{depth}(\blk b) = \texttt{depth}(\blk a) + 6$, the settlement will take about one minute.
Invalidation of the payment, e.\,g.\ due to a double spend, implies a fork of the subblock tree of length 6 or more.
Tailstorm discounts mining rewards according to the depth of the subblock tree; whereas the tree could have achieved depth 60, it now can achieve depth at most 54.
One tenth of the minted rewards is lost, and the cost of the coffee is dwarfed.

\egroup 

\subsection{Tailstorm Prototype}

We have implemented Tailstorm and make the code available online~\anoncite{clientImpl}.
We started from a fork Bitcoin Unlimited's implementation of Bitcoin Cash.
Our fork implements the Tailstorm consensus layer and incentive mechanism as specified in Section~\ref{sec:spec}.
The prototype uses $k=3$, but is easily configurable by changing a compile-time flag.
The DAA is calibrated for a target summary block interval of ten minutes.
Subblocks are expected to arrive every 200 seconds.

To minimize propagation delays, we have implemented several network compression mechanisms.
First, we adapt the Graphene protocol~\cite{ozisik2019GrapheneEfficient} to avoid redundant transmission of transactions.
Second, we adapt the Compact Block protocol~\cite{Corallo:2016} to reduce the size of summary blocks.
Third, the transaction list is encoded bit-efficiently in each subblock.

We leave the implementation of Tailstorm's application layer, i.\,e.\ the transaction handling as described in Section~\ref{sec:transaction}, for future work.
Finally, we note that our implementation has minimal testing and, being a prototype, is not ready for production use.

\section{Discussion}
\label{sec:discussion}

Tailstorm inherits the consistency guarantees of parallel PoW consensus~\cite{keller2022ParallelProofofwork}.
To increase fairness, we discount rewards based on the tree structure of subblocks which itself is inspired from Bobtail~\cite{bissias2020BobtailImproved}.
To obtain fast confirmations, we write transactions into subblocks like Storm~\cite{awemany2019StormWhitepaper}.
Alongside these foundational influences, we incorporate insights from a variety of related works, which we here can only list partially.
For more details we refer to the surveys of Garay and Kiayias~\cite{garay2020SoKConsensus}, describing the different ways of defining the consensus problem, and Bano et al.~\cite{bano2019SoKConsensus}, focusing on the different solution approaches and protocols.

Tailstorm improves fairness by avoiding orphans and discounting rewards.
The former is not new~\cite{amoressesar2021GeneralizingWeighted, sompolinsky2021PhantomGhostdag, sompolinsky2015SecureHighrate}, however, the motivation differs.
Sompolinsky and Zohar~\cite{sompolinsky2015SecureHighrate} propose to improve Bitcoin's \emph{consistency} by referencing blocks that would otherwise be orphaned (called \emph{uncles}) and then counting both blocks and uncles (GHOST-rule) whereas Bitcoin counts only the blocks (longest chain rule).
However, the authors explicitly avoid rewarding uncles and hence do not improve \emph{fairness}.
Follow-up work~\cite{amoressesar2021GeneralizingWeighted, sompolinsky2021PhantomGhostdag} does not discuss rewards at all.
In Ethereum PoW, a deployed variant of GHOST~\cite{sompolinsky2015SecureHighrate}, and Kaspa, a deployed variant of GHOSTDAG~\cite{sompolinsky2021PhantomGhostdag}, uncles receive \emph{partial} rewards and the blocks that include uncles get \emph{additional} rewards.
The unfair dynamic of Bitcoin, where included blocks receive the full reward and orphans  none, is largely preserved.
We think that GHOST-like protocols can become more fair by applying Tailstorm's ideas: discounting of rewards, applied equally to all involved miners.

Pass and Shi~\cite{pass2017FruitChainsFair} explicitly set out to increase fairness in PoW.
Their Fruitchains protocol uses two kinds of blocks like Tailstorm.
Blocks record fruits and fruits record transactions.
Unlike in Tailstorm, both fruits and blocks require PoW.
Applying the \emph{2-for-1 trick} of Garay et al.~\cite{garay2015BitcoinBackbone}, both are mined with the same hash puzzle observing different bits of the output.
Fruits have a lower difficulty and thus are created more frequently.
The rewards are distributed evenly across recent fruits while blocks receive nothing.
According to Zhang and Preneel~\cite{zhang2019LayCommon}, Fruitchains are more vulnerable to incentive layer attacks than Bitcoin for block race advantage~$\gamma = 0$, and thereby less fair.
Furthermore, Fruitchains suffer from a tradeoff between fairness and transaction confirmation time~\cite{zhang2019LayCommon}.
A related line of research~\cite{bagaria2019PrismDeconstructing, yu2020OHIEBlockchain} extends the \emph{2-for-1} into an \emph{n-for-1} trick and makes the miners work on multiple chains in parallel.
The chains are then interleaved to form a single coherent transaction ledger.
Unlike Fruitchains, however, Prism~\cite{bagaria2019PrismDeconstructing} and OHIE~\cite{yu2020OHIEBlockchain} focus on consistency, liveness, and throughput.
We leave for future work to investigate whether and how the fairness of these protocols can be improved by applying Tailstorm's reward discounting.

Kiffer and Rajaraman~\cite{kiffer2021HaPPYmineDesigning} propose to discount rewards inversely proportional to the overall hash rate participating in the system.
This reduces centralization in their model, but we worry that motivating lower hash rates might harm the primary goals of consensus: consistency and liveness.
Their mechanism poses an alternative to Bitcoin's halving mechanism, which statically reduces block rewards by 50\,\% roughly once every 4 years, to avoid long term inflation of the cryptocurrency.
In contrast, Tailstorm's reward discounting is focused on the short term and punishes non-linearities within a single subblock tree.

But our approach is not without limitations.
In Section~\ref{sec:fairness}, we analyse the impact of message propagation delays on fairness.
We assume that these delays are uniform, affecting all miners equally.
However, in practice, propagation delays depend on the connectivity of individual miners.
Some inequalities arise from economies of scale, larger miners can invest more in low latency connections,
others from the underlying physics of information propagation:
miners located in the same region or on the same continent implicitly form are cartel, whereas joining the network from a distant location puts them at a disadvantage.
We leave for future work to analyze how Tailstorm is affected by more realistic network topologies~\cite{alzayat2021ModelingCoordinated, delgadosegura2019TxProbeDiscovering, mariem2020AllThat, rohrer2019KadcastStructured, rohrer2021BlockchainLayer}.

In Section~\ref{sec:reinforcement_learning}, we perform a search for optimal attack strategies.
Our search algorithm is based on reinforcement learning (RL), which means it is not exhaustive.
It is possible that the RL agent did not discover the optimal strategy against certain protocols.
This challenges our conclusion that Tailstorm is less susceptible to incentive layer attacks than the other protocols, Bitcoin and $\mathcal B_k$.
To encourage further exploration we release the RL Gym environment on PyPI~\anoncite{cpr:pypi}, allowing others to discover more effective attacks.
Assuming the absence of better strategies, our results can be confirmed by encoding our attack space as a Markov Decision Process (MDP) and employing exhaustive search techniques, as has been done for Bitcoin~\cite{gervais2016SecurityPerformance, sapirshtein2016OptimalSelfish}, Ethereum~\cite{barzur2020EfficientMDP} and other longest chain protocols~\cite{zhang2019LayCommon}.
However, to the best of our knowledge,  such techniques have not yet been successfully applied to DAG-based protocols like Tailstorm.

Another limitation is that our action space might not cover all possible attacks.
We closely follow the modeling of the selfish mining attack space against Bitcoin~\cite{sapirshtein2016OptimalSelfish}, where we observe consent that the four actions \texttt{Adopt}, \texttt{Match}, \texttt{Override}, and \texttt{Wait} are indeed complete.
We are convinced that adding two actions for summary formation, \texttt{Inclusive} and \texttt{Exclusive}, is enough to represent the most profitable attacks.
However, our conclusions can be challenged by presenting a more effective incentive layer attack against Tailstorm, which cannot be expressed in our attack space.
We encourage this endeavour by making all analytical code available online~\anoncite{cpr:repo}.
After modifying the attack space, new strategies can be drafted, evaluated, and optimized, while the virtual environment, protocol specifications, and RL attack search can be reused without changes.

We demonstrate in Section~\ref{sec:reinforcement_learning} that ignoring subblocks is no major concern in the long term, i.\,e., after the mining difficulty was adjusted to the attack.
But, as we point out in Appendix~\ref{apx:short_term_exlusive}, there is a small incentive to do so in the short term (before difficulty adjustment).
This stands in contrast to Bitcoin, where selfish mining implies negative outcome in the short term and positive outcome in the long term.
We feel this tradeoff is favorable to Tailstorm.

On a separate note, Carlsten et al.~\cite{carlsten2016InstabilityBitcoin} demonstrate that selfish mining becomes more profitable when considering transaction fees in addition to mining rewards.
They present a strategy targeting Bitcoin which leverages transaction fees to outperform honest behavior for any $\alpha > 0$ and $\gamma < 1$.
Similar attacks are likely feasible against all PoW cryptocurrencies, including Tailstorm, however they also exceed the scope of this paper.

Lastly, Arnosti and Weinberg~\cite{arnosti2022BitcoinNatural} show that even small cost imbalances or economies of scale lead to highly centralized PoW mining ecosystems.
While, Tailstorm reduces the imbalances compared to other PoW protocols we studied, it cannot fully mitigate centralization.
Our notion of fairness revolves around rewarding miners in proportion to their hash rate.
This definition is not new~\cite{alzayat2021ModelingCoordinated, birmpas2020FairnessEfficiency, kiayias2016BlockchainMining, pass2017FruitChainsFair, zhang2019LayCommon}.
However, from an economic standpoint, to prevent centralization, miners should be rewarded in proportion to their operational mining costs~\cite{arnosti2022BitcoinNatural}.
Unfortunately, achieving this goal seems to be impossible.
Mining costs primarily depend on the price and the availability of energy and specialized mining hardware.
Purchasing energy in large quantities tends to be more cost-effective, and strong miners may even consider operating their own power plants.
Scaling up operations also enables miners to develop and deploy more efficient mining hardware, and then selling it to weak miners after it has become outdated~\cite{yaish2023PricingASICs}.
Given that these unfair scaling effects do not only affect Tailstorm, but PoW cryptocurrencies in general, one might be lead to consider alternative approaches like proof-of-stake~\cite{gilad2017AlgorandScaling, kiayias2017OuroborosProvably, sliwinski2021AsynchronousProofofstake}.
But even then, it remains questionable whether any permissionless system can fully avoid centralization~\cite{kwon2019ImpossibilityFull}.

\section{Conclusion}\label{sec:conclusion}

Tailstorm integrates parallel PoW~\cite{keller2022ParallelProofofwork}, partial transaction confirmation~\cite{awemany2019StormWhitepaper,bissias2020BobtailImproved}, and a novel incentive mechanism---reward discounting---into a PoW cryptocurrency that provides fast and secure confirmations for its users and fair rewards for its miners.
We thereby solve a long standing issue of longest chain protocols, where the operator has to choose between either a short block interval with fast but less reliable confirmations and unfair rewards, or a long block interval with slow confirmations and more fair but infrequent rewards.
Our prototype implementation demonstrates that Tailstorm can serve as a replacement for Bitcoin not only theoretically but also in practice.

\ifdefined\anon\else
{
  \small

  \section*{Contribution Statement}
  George originally developed the idea of reward discounting as a way to combat withholding attacks in Bobtail.
  George and Gregory refined that idea into an early version of Tailstorm suitable for a hard fork of Bitcoin Cash.
  They also contributed to the prototype implementation.
  Patrik contributed the protocol specification, simulation-based analyses, and reinforcement learning interface.
  George contributed the analytical orphan rate analysis.
  George and Patrik contributed the hard-coded reference policies.
  Ben and Patrik conducted the attack search with reinforcement learning.
  Guided by earlier stages of the analyses, Patrik provided improvements to Tailstorm consensus and its transaction handling.
  Patrik and George wrote this paper.

  \section*{Acknowledgements}
  We wish to thank Bitcoin Unlimited for their financial and technical support as well as Michael Fröwis for his review of this work and for the helpful suggestions he provided.
}
\fi 

\bibliography{references, patrik-generated}

\appendix

\section{Reference Protocols}\label{apx:protocols}

To enable comparison of Tailstorm to Bitcoin~\cite{nakamoto2008BitcoinPeertopeer} and $\mathcal B_k$~\cite{keller2022ParallelProofofwork}, we specify these protocols using the system model described in Section~\ref{sec:model}.

\subsection{Bitcoin}\label{apx:bitcoin}

We specify the Bitcoin protocol in Algorithm~\ref{alg:nakamoto}.
Bitcoin's chain structure is simple: each block has a PoW and exactly one parent; the $\texttt{height}$ property tracks the number of ancestors.
Protocol-compliant nodes try to extend the longest chain of blocks.
Tie-breaking between multiple blocks of the same height happens in favour of the block first seen.
Freshly mined blocks are shared immediately.

\begin{algorithm}[t]
  \caption{Bitcoin: Chain Structure and Node}
  \label{alg:nakamoto}

  \Fn{\texttt{Root}()}{
    \Return block template $b$ with $\texttt{height}(b) = 0$\;
  }

  \BlankLine
  \Fn{\texttt{Validate}({\normalfont block} $b$)}{
      $p \gets$ first parent of $b$\;
      \Return $\texttt{pow}(b) \land (|\texttt{parents}(b)| = 1) \land (\texttt{height}(b) = \texttt{height}(p) + 1)$\;
  }

  \BlankLine
  \Fn{\texttt{Update}({\normalfont tip} $t$, {\normalfont block} $b$)}{
    \lIf{$\texttt{height}(b) > \texttt{height}(t)$}{%
      \KwRet{$b, [b], []$}%
    }
    \lElse{%
      \KwRet{$t, [], []$}%
    }
  }

  \BlankLine
  \Fn{\texttt{Extend}({\normalfont tip} $t$)}{
    \Return block template $b$ with
    $\texttt{parents}(b) = [t]$\, and
    $\texttt{height}(b) = \texttt{height}(t) + 1$\;
  }
\end{algorithm}

For dishonest behaviour we closely model the attack space used by Sapirshtein et al.~\cite{sapirshtein2016OptimalSelfish}.
Recall Tailstorm's attack space from Section~\ref{sec:attacks}.
For Bitcoin, we restrict the observations to $(h_\text a, h_\text d)$ (comp. Sec.~\ref{sec:attacks:obs}) and use only the first half of the action tuple, \emph{withhold}, with the actions \texttt{Wait}, \texttt{Match}, \texttt{Override}, and \texttt{Adopt} (comp. Sec.~\ref{sec:attacks:actions}).
Algorithm~\ref{alg:bitcoin:attacker} implements the dishonest node (comp. Alg.~\ref{alg:ts:attacker}).

\begin{algorithm}[t]
  \caption{Bitcoin: Attacker}
  \label{alg:bitcoin:attacker}

  \BlankLine
  \Fn{$\texttt{Update}(\text{tip } t, \text{block } b)$}{
    $\emph{pref}, \emph{share} \gets t, []$\;
    \lIf{$b$ is own block}{$\emph{pref} \gets b$}
    $W \gets$ set of withheld blocks (= blocks not shared before)\;
    $\emph{withhold} \gets \mathcal P(h_\text a, h_\text d)$\;
    \Switch{\emph{withhold}}{
      \lCase{\texttt{Adopt}}{%
        $\emph{pref} \gets b_\text{d}$%
      }
      \lCase{\texttt{Match}}{%
        $\emph{share} \gets \left\{ x \in W \mid \texttt{height}(x) \leq \texttt{height}(b_\text{d}) \right\}$%
      }

      \lCase{\texttt{Override}}{%
        $\emph{share} \gets \left\{ x \in W \mid \texttt{height}(x) \leq \texttt{height}(b_\text{d}) + 1 \right\}$%
      }
      \lCase{\texttt{Wait}}{%
        nothing%
      }
    }
    \Return \emph{pref}, \emph{share}, $[]$\;
  }
\end{algorithm}

We define a single reference policy for Bitcoin, SM1. It was originally proposed by Eyal and Sirer~\cite{eyal2014MajorityNot}, but we here use the more formal definition by Sapirshtein et al.~\cite{sapirshtein2016OptimalSelfish}.

\begin{description}
  \item[SM1]
    Withhold blocks as long as the honest chain is shorter:\par
    \texttt{Adopt} if $h_\text{d} > h_\text{a}$.
    \texttt{Match} if $h_\text{d} = h_\text{a} = 1$.
    \texttt{Override} if $h_\texttt{d} = h_\text{a} - 1$ and  $h_\texttt{d} \geq 1$.
    Otherwise \texttt{Wait}.
\end{description}

\subsection{$\mathcal B_k$}\label{apx:bk}

We specify the $\mathcal B_k$ protocol in Algorithm~\ref{alg:bk}.
$\mathcal B_k$ is similar to Tailstorm in several ways.
Both protocols only require proofs-of-work for subblocks, and constructing a new summary necessitates $k$ subblocks that confirm the previous summary.
Compliant nodes extend the longest chain of summaries and break ties by the number of confirming subblocks.
However, there are two significant differences between the two protocols.
First, $\mathcal B_k$ subblocks do not form a tree but instead directly reference the previous summary.
This makes it pointless to track the subblock depth, and the discount reward scheme cannot be applied.
Second, $\mathcal B_k$ employs a leader election mechanism to restrict who can append summaries based on the hash values of the subblocks.
Specifically, lines~\ref{l:bk:leader:a}, \ref{l:bk:leader:b}, and~\ref{l:bk:leader:c} in Algorithm~\ref{alg:bk} ensure that only the miner of the smallest subblock can add the next summary.
We have made minor changes to the terminology used by Keller and Böhme~\cite{keller2022ParallelProofofwork} to align with Tailstorm: what they refer to as a \emph{block}, we call a summary, and what they refer to as a \emph{vote}, we call a subblock.

Tailstorm's attack space, as defined in Section~\ref{sec:attacks}, translates directly to $\mathcal B_k$.
The only modification is that we omit the irrelevant variables $d_\text a$, $d_\text a'$, and $d_\text d$ for the subblock tree depth from the observation space.
The reference policies presented in Section~\ref{sec:attacks:policies} can be applied without modification.

\begin{algorithm}[t]
  \caption{$\mathcal B_k$: Chain Structure and Node}
  \label{alg:bk}

  \Fn{\texttt{Root}()}{
    \Return block template $b$ with $\texttt{summary}(b) = \texttt{true}$ and
    $\texttt{height}(b) = 0$\;
  }

  \BlankLine
  \Fn{$\texttt{Validate}(\text{block } b)$}{
    \uIf{$\texttt{summary}(b)$}{
      $p \gets$ last summary before $b$\;
      $S \gets$ subblocks between $b$ and $p$\;
      $l \gets$ subblock with lowest hash value in $S$\label{l:bk:leader:a}\;
      \Return $
      |S| = k \land
      \texttt{height}(b) = \texttt{height}(p) + 1 \land
      \texttt{miner}(b) = \texttt{miner}(l)
      $\label{l:bk:leader:b}\;
    }
    \Else(\tcc*[f]{$b$ is subblock}){
      $p \gets$ first parent of $b$\;
      \Return $
      \texttt{pow}(b) \land
      |\texttt{parents}(b)| = 1 \land
      \texttt{height}(b) = \texttt{height}(p)$\;
    }
  }

  \BlankLine
  \Fn{$\texttt{Preference}(\text{summary } s, \text{summary } b)$}{
    \emph{hs}, \emph{hb} $\gets \texttt{height}(s), \texttt{height}(b)$\;
    \emph{ns}, \emph{nb} $\gets$ number of subblocks after $(s,b)$\;
    \emph{vs}, \emph{vb} $\gets$ smallest hash value in $(\texttt{parents}(s),\texttt{parents}(b))$\;
    \Return $
    (hb > hs) \lor
    (hb = hs \land nb > ns) \lor
    (hb = hs \land nb = ns \land vb < vs)
    $\label{l:bk:leader:c}\;
  }

  \BlankLine
  \Fn{$\texttt{Update}(\text{tip } t, \text{block } b)$}{
    $\emph{pref}, \emph{share}, \emph{append} \gets t, [b], []$\;
    \uIf{$\texttt{summary}(b)$}{
      \lIf{$\texttt{Preference}(\emph{pref},b)$}{$\emph{pref} \gets b$}
    }
    \Else(\tcc*[f]{$b$ is subblock}){
      $p \gets$ last summary before $b$\;
      \lIf{$\texttt{Preference}(\emph{pref},p)$}{$\emph{pref} \gets p$}
      $S \gets$ all subblocks in subblock tree confirming $p$\;
      \If{$|S| \geq k$}{
        $R \gets$ select $k$ subblocks from $S$\;
        create summary template $s$ with $\texttt{parents}(s) = R$\;
        \emph{append} $\gets [s]$\;
      }
    }
    \Return $t$, \emph{share}, \emph{append}\;
  }

  \BlankLine
  \Fn{\texttt{Extend}({\normalfont tip} $t$)}{
    \Return block template $b$ with\\
    $\quad\texttt{summary}(b) = \texttt{false}$,
    $\texttt{parents}(b) = [t]$, and
    $\texttt{height}(b) = \texttt{height}(t)$\;
  }
\end{algorithm}

%
%

\section{Ignoring Subblocks to Maximize Rewards in the Short Term}\label{apx:short_term_exlusive}

The \texttt{Exclusive} action of our attack space defined in Section~\ref{sec:attacks} enables attackers to ignore foreign subblocks that lie off the main branch in the subblock tree.
Including such subblocks in the next summary would reduce the depth of the tree and, due to Tailstorm's discounting mechanism, also the individual subblock rewards.
In Section~\ref{sec:reinforcement_learning}, we search for attack strategies maximizing the \emph{normalized reward} metric, which captures rewards under the hypothesis that the difficulty was already adjusted to the increased number of orphans caused by the attack.
The results in Section~\ref{sec:reinforcement_learning} demonstrate that ignoring subblocks is no major concern in the long term, that is, after the difficulty adjustment algorithm (DAA) has stabilized.
In this section, we investigate the profitability of ignoring subblocks in the short term, i.\,e., before the DAA modifies the subblock mining rate~$\lambda$.

Let $\bar d_\text{pow} = \nicefrac1\lambda$ denote the expected time required to mine a subblock.
Suppose that $k$ subblocks are mined, all confirming the same parent summary block~$P$.
All are connected in a chain except one, which we label $o$.
Miners have two options.
They can either assemble a new summary block $Q$ from the $k$ subblocks now, allowing them to start mining subblocks with parent $Q$, or they can continue to mine with parent $P$, hoping to achieve a fully linear chain of subblocks with higher rewards.

We observe both options from time $t_0$, when parent summary $P$ has been proposed and no confirming subblocks have been mined, and time $t_{k+1}$, when the next $k+1$ subblocks are available. Time $t_k$ is the arrival time of the $k$th subblock, i.\,e., the time when assembling the new summary~$Q$ first becomes possible.
Following the definition of our mining process in Section~\ref{sec:model}, the expected values for the times $\{t_i\}_{i \geq 0}$ are $\{i \cdot \bar d_\text{pow}\}$.

Taking the first option, assembling $Q$ at time $t_k$, summary $Q$ is of depth $k - 1$ and, according to the discount rule in Formula~\ref{f:discount}, assigns $c / k \cdot (k-1)$ units of rewards to each of the $k$ subblocks.
The scaling factor $c$ has to be chosen by the operator; to simplify our argument, we assume $c = 1$.
In total, summary $Q$ allocates $k - 1$ units of reward.
The miners receive $(k - 1) / ( k \cdot \bar{d}_\text{pow} )$ units of reward per unit of time.

The second option, continue mining with parent $P$, \emph{can} result in a linear chain of $k$ subblocks at time $t_{k + 1}$.
This would allow for a summary that allocates $k$ units of reward.
The miners would receive $k / ( (k+1) \cdot \bar{d}_\text{pow} )$ units of reward per time.

Option two, delaying the summary and ignoring one subblock off the main branch gives $1 / (k^2 - 1)$ more units of reward per time than option one, including all subblocks and summarizing early.
So it seems there is a small incentive to act selfishly in the short term.
This stands in contrast to Bitcoin, where selfish mining implies strictly negative expected utility in the short term and positive utility in the long term.

We note though, that this result is preliminary.
We leave for future work to remove the assumption that subblock $k+1$ is indeed of depth $k$, and to address the remaining cases with more than one subblock off the main chain.
Additionally, it would be interesting to conduct the analysis from the viewpoint of an individual miner, instead of the collective of all miners like we do above.

Lastly, we think that there is enough social pressure to follow the rules.
Mining is typically a public endeavour: miners tend to identify themselves, particularly those operating in pools.
When a pool operator chooses to ignore subblocks, this will be apparent to all miners and they will join a different pool.
Conversely, if the majority of miners agrees on maximizing short term reward, they might as well change the protocol arbitrarily, e.\,g., by simply scaling up the mining rewards.

\end{document}